\documentclass[useAMS,usenatbib]{mn2e}
\usepackage{graphics}
\usepackage{graphicx}
\usepackage{psfig}
\input{epsf}
\def\half{{1\over 2}}

\def\etal{{\it et~al.\ }}
\def\eg{{\it e.g.,~}}
\def\ie{{\it i.e.~}}

\def\ltsima{$\; \buildrel < \over \sim \;$}
\def\simlt{\lower.5ex\hbox{\ltsima}}
\def\gtsima{$\; \buildrel > \over \sim \;$}
\def\simgt{\lower.5ex\hbox{\gtsima}}

\def \lleq {\lower0.9ex\hbox{ $\buildrel < \over \sim$} ~}
\def \ggeq {\lower0.9ex\hbox{ $\buildrel > \over \sim$} ~}

\def \l    {\Lambda}
\def\rhoc{\rho_{0{\rm c}}}

\def \omx  {\Omega_X}
\def \om   {\Omega_m}
\def \omt  {\Omega_{0 {\rm m}}}
\def \ol   {\Omega_{\Lambda}}

\def \beq  {\begin{equation}}
\def \eeq  {\end{equation}}
\def \ber  {\begin{eqnarray}}
\def \eer  {\end{eqnarray}}

\def\apj{{Astroph.\@ J.\ }}
\def\mn{{Mon.\@ Not.\@ Roy.\@ Ast.\@ Soc.\ }}
\def\asta{{Astron.\@ Astrophys.\ }}
\def\aj{{Astron.\@ J.\ }}
\def\prl{{Phys.\@ Rev.\@ Lett.\ }}
\def\pd{{Phys.\@ Rev.\@ D\ }}
\def\nucp{{Nucl.\@ Phys.\ }}

\def\plb {{Phys.\@ Lett.\@ B\ }}
\def \jetpl {JETP Lett.\ }

\begin{document}

\title{Is there Supernova Evidence for  Dark Energy Metamorphosis ?}

\author[Ujjaini Alam, Varun Sahni, Tarun Deep Saini, A. A. Starobinsky]{Ujjaini Alam$^{1,4}$, Varun Sahni$^{1,5}$, Tarun Deep Saini$^{2,6}$ and A. A. Starobinsky$^{3,7}$ \\
  $^{1}$ Inter University Centre for Astronomy $\&$ Astrophysics, Pune, India\\
  $^{2}$ Institute of Astronomy, Madingley Road, Cambridge, UK \\
  $^{3}$ Landau Institute for Theoretical Physics, Kosygina 2, Moscow
  119334, Russia\\
  $^{4}$ujjaini@iucaa.ernet.in \\
  $^{5}$varun@iucaa.ernet.in \\
  $^{6}$tarun@ast.cam.ac.uk \\
  $^{7}$alstar@landau.ac.ru \\
  } \maketitle


\begin{abstract}
We reconstruct the equation of state $w(z)$ of dark energy (DE) using
a recently released data set containing 172 type Ia supernovae without
assuming the prior $w(z) \geq -1$ (in contrast to previous studies).
We find that dark energy evolves rapidly and metamorphoses from dust-like
behaviour at high $z$ ($w \simeq 0$ at $z \sim 1$) to a strongly
negative equation of state at present ($w \lleq -1$ at $z \simeq 0$).
Dark energy metamorphosis appears to be a robust phenomenon which
manifests for a large variety of SNe data samples provided one does not
invoke the weak energy prior $\rho + p \geq 0$.  Invoking this prior
considerably weakens the rate of growth of $w(z)$.  These results
demonstrate that dark energy with an evolving equation of state
provides a compelling alternative to a cosmological constant if data
are analysed in a prior-free manner and the weak energy condition is
not imposed by hand.
\end{abstract}

\section{Introduction }
One of the most tantalizing observational discoveries of the past
decade has been that the expansion of the universe is speeding up
rather than slowing down.  An accelerating universe is strongly
suggested by observations of type Ia high redshift supernovae provided
these behave as standard candles.  The case for an accelerating
universe is further strengthened by the discovery of Cosmic Microwave
Background (CMB) anisotropies on degree scales (which indicate
$\Omega_{\rm total} \simeq 1$) combined with a low value for the
density in clustered matter $\om \simeq 1/3$ deduced from galaxy
redshift surveys.  All three sets of observations strongly suggest
that the universe is permeated by a relatively smooth distribution of
`dark energy' (DE) which dominates the density of the universe
($\Omega_{\rm DE} \simeq 2\om \simeq 2/3$) and whose energy momentum
tensor violates the strong energy condition ($\rho + 3p \geq 0$) so
that $w_{\rm DE} = p/\rho < -1/3$.

Although a cosmological constant ($w = -1$) provides a plausible
answer to the conundrum posed by dark energy, it is well known that
the unevolving cosmological constant faces serious `fine tuning'
problems since the ratio between $\rho_\l = \l/8\pi G$ and the
radiation density, $\rho_r$, is already a miniscule $\rho_\l/\rho_r
\sim 10^{-54}$ at the electroweak scale ($T \sim 100$ GeV) and even
smaller, $\rho_\l/\rho_r \sim 10^{-123}$, at the Planck scale ($T \sim
10^{19}$ GeV).  This issue is further exacerbated by the `cosmological
constant problem' which arises because the $\l$-term generated by
quantum effects is enormously large $\rho_\l \ggeq m_{\rm Pl}^4$,
where $m_{\rm Pl} \simeq 1.2 \times 10^{19}$ GeV is the Planck mass
\citep{zel68,wein89}.

Although the cosmological constant problem remains unresolved, the
issue of fine tuning which plagues $\l$ has led theorists to explore
alternative avenues for DE model building in which either DE or its
equation of state are functions of time.  (Following \citet{sahni03}
we shall refer to the former as Quiessence and to the latter as
Kinessence.)  Inspired by inflation, the first dark energy models were
constructed around a minimally coupled scalar field (quintessence)
whose equation of state was a function of time and whose density
dropped from a large initial value to the small values which are
observed today \citep{peebles88,wetterich88}.  (`Tracker' quintessence
models had the advantage of allowing the current accelerating epoch to
be reached from a large family of initial conditions \citep{cald98}.)

Half a decade after SNe-based observations pointed to the possibility
that we may be living in an accelerating universe, the theoretical
landscape concerning dark energy has evolved considerably \citep[see
the reviews][]{ss00,carroll01,pr02,sahni02a,paddy03}.  In addition to
the cosmological constant and quintessence, the current paradigm for
DE includes the following interesting possibilities:

\begin{itemize}

\item {\bf Dark energy with $w \leq -1$}
\citep{chiba00,caldwell,innes,ss02a,as02,caldwell03,carroll03,frampton03a,frampton03b,singh,johri}

\item {\bf The Chaplygin Gas} whose equation of state drops from
$w=0$ at high redshifts to $w \simeq -1$ today \citep{chap1}

\item {\bf Braneworld models} in which the source for cosmic
acceleration rests in the {\em gravity sector} rather than in the
matter sector of the theory \citep{DDG,ss02a,maeda}

\item {\bf Dark energy models with negative potentials}
\citep{felder02,kallosh02,alam03}

\item {\bf Interacting models of dark matter and dark energy}
\citep{amendola00,chimento03,hoffman03}

\item{\bf Modified gravity and scalar-tensor theories}
\citep{beps00,bm00,bp00,damour02}

\item{\bf Dark energy driven by quantum effects}
\citep{sh98,parker99}

\item{\bf Dark energy with a late-time transition in the equation of state}
\citep{bassett,corasaniti}

\item {\bf Unified models of dark energy and inflation}
\citep{pv99,copeland01,sahni02c} etc.

\end{itemize}

Faced with the current plethora of dark energy scenarios the
concerned cosmologist is faced with two options:

(i) She can test {\em every single} model against observations,

(ii) She can take a more flexible approach and determine the
properties of dark energy in a {\em model independent manner}.

In this paper we proceed along route (ii) and demonstrate that model
independent reconstruction brings us face to face with exciting new
properties of dark energy.

Applying the techniques developed in \cite{saini00,sahni03} to a new
data set consisting of $172$ Supernovae from \cite{tonry03} and an
additional 22 Supernovae from \cite{hzt_new} we show that the DE
equation of state which best fits the data evolves from $w \simeq 0$
at $z \simeq 1$ to $ -1.2 \lleq w \lleq -1$ today.  {\em An evolving
  equation of state of DE is favoured by the data over a cosmological
  constant for a large region in parameter space.}

\section{\bf Model independent reconstruction of dark energy}

Supernova observations during the previous decade have been pioneered
by two teams: The High-z Supernova Search Team (HZT) \citep{riess} and
the Supernova Cosmology Project (SCP) \citep{perl}. The enormous
efforts made by these two teams have changed the way cosmologists view
their universe.  A recent analysis \citep{tonry03} of 172 type Ia
supernovae by HZT gives the following bounds on the cosmic equation of
state (at $95\%$ CL)
\beq
-1.48 < w < -0.72,
\label{eq:state0}
\eeq
when the 2dFGRS prior $\om h = 0.2 \pm .03$ is assumed \citep{2dF}.
A similar bound
\beq
-1.61 < w < -0.78,
\label{eq:state1}
\eeq
is obtained for a new sample of high-z supernovae by SCP \citep{knop}.
\footnote{It is interesting that, when no priors are set on $\om$, the
dark energy equation of state becomes virtually unbounded from below
and has a $99\%$ confidence limit of being $< -1$ ! \citep{knop}}

These results clearly rule out several DE contenders including a
tangled network of cosmic strings ($w \simeq -1/3$) and domain walls
($w \simeq -2/3$). However a note of caution must be added before we
apply (\ref{eq:state0}) or (\ref{eq:state1}) to the wider class of DE
models discussed in the introduction.  Impressive as the bounds in
(\ref{eq:state0}) \& (\ref{eq:state1}) are, they strictly apply only
to dark energy having a {\em constant equation of state} since this
prior was assumed both in the analysis of the supernova data set as
well as in the 2dFGRS study \citep{tonry03,knop}.  Aside from the
cosmological constant ($w = -1$), the topological defect models
alluded to earlier and the sine-hyperbolic scalar field potential
\citep{ss00,um00,sahni03} no viable DE models exist with the property
$w = {\rm constant}$.  Indeed, most models of dark energy
(Quintessence, Chaplygin gas, Braneworlds, etc.) can show significant
evolution in $w(z)$ over sufficiently large look back times.

In this paper we shall reconstruct the properties of dark energy {\em
  without assuming any priors} on the cosmic equation of state.  (The
dangers of imposing priors on $w(z)$ have been highlighted in
\cite{maor02} and several of our subsequent results will lend support
to the conclusions reached in this paper.)

\subsection{Cosmological reconstruction of $w(z)$}

Cosmological reconstruction is based on the observation that, in a
spatially flat universe, the luminosity distance and the Hubble
parameter are related through the equation \citep{st00,turner,nak99}:
\beq\label{eq:H}
H(z)=\left[{d\over dz}\left({d_L(z)\over 1+z}\right)\right]^{-1}~.
\eeq
Thus knowing $d_L$ we can unambiguously determine the Hubble parameter
as a function of the cosmological redshift.  Next, the Einstein
equations
\ber
H^2 &\simeq& \frac{8\pi G}{3}\lbrack \rho_m + \rho_{\rm DE} \rbrack~,
\nonumber\\
q &=& -\frac{\ddot a}{aH^2} = \frac{4\pi G}{3 H^2}\sum_i (\rho_i + 3p_i)~,
\label{eq:einstein}
\eer
are used to determine the energy density and pressure of dark energy:
\ber\label{eq:energy}
\rho_{\rm DE} &=& \rho_{\rm critical} - \rho_{\rm m} =
\frac{3H^2}{8\pi G}(1 - \Omega_{\rm m}(x))~,\nonumber\\
p_{\rm DE} &=& \frac{H^2}{4\pi G}(q - \frac{1}{2})~,
\eer
where $\rho_{\rm critical} = 3H^2/8\pi G$ is the critical density of a
FRW universe. The equation of state of DE $w_{\rm eff} = p_{\rm
  DE}/\rho_{\rm DE}$ follows immediately \citep{saini00}
\beq\label{eq:state}
w_{\rm eff}(x) = {2 q(x) - 1 \over 3 \left( 1 - \Omega_{\rm m}(x) \right)}
\equiv \frac{(2 x /3) \ d({\rm ln}H)/dx - 1}{1 \ - \ (H_0/H)^2
\omt \ x^3}\,\,,
\eeq
where $\omt = 8\pi G\rho_{0 {\rm m}}/3H_0^2$, $x = 1+z$.
In quintessence models and in $\l$CDM, the equation (\ref{eq:state})
determines the true `physical' equation of state of dark energy.
However the subscript `eff' in $w_{\rm eff}$ stresses the fact that
this quantity should be interpreted as an `effective' equation of
state in DE models in which gravity is non-Einsteinian or in models in
which dark energy and dark matter interact.  Examples of the former
include Braneworld models and scalar-tensor theories.  It is well
known that in a large class of Braneworld models the Hubble parameter
does not adhere to the Einsteinian prescription (\ref{eq:einstein})
since it includes explicit interaction terms between dark matter and
dark energy \citep{DDG,ss02a}.  In this case the equation of state
determined using (\ref{eq:state}) can still be used to characterize
DE, but physical interpretations of $w_{\rm eff}$ need to be treated
with caution.  \footnote{One way around this difficulty is to define
  observables solely in terms of $H$ and its derivatives \citep[called
  `Statefinders' in ][] {sahni03}.  A detailed discussion of these
  issues can be found in \cite{alam03a}.}

One route towards the meaningful reconstruction of $w(z)$ lies in inventing
a sufficiently versatile fitting function for either $d_L(z)$ or $H(z)$.
The parameters of this fitting function are determined by matching to
Supernova observations and $w(z)$ is determined from (\ref{eq:H})
and (\ref{eq:state}).\footnote{Alternatively one could apply an ansatz to
$w(z)$ itself \citep{chiba,albrecht,copeland,efstathiou,maor02,linder}.
See \cite{alam03a} for a summary of different approaches to cosmological
reconstruction. Non-parametric approaches are discussed in
\cite{wang01,hut03,saini03}; see also \cite{daly03,nunes}.}
Our reconstruction exercise will be based upon the following flexible and
model independent ansatz for the Hubble parameter \citep{sahni03}
\beq
H(x) =  H_0\left\lbrack \om x^3 + A_0 + A_1x + A_2 x^2\right\rbrack^\half\,\, ,
\label{eq:taylor}
\eeq
where $x= 1+z$. This ansatz for $H(z)$ is exact for the cosmological
constant $w = -1$ ($A_1 = A_2 = 0$) and for DE models with $w = -2/3$
($A_0 = A_2 = 0$) and $w = -1/3$ ($A_0 = A_1 = 0$).  It has also been
found to give excellent results for DE models in which the equation of
state varies with time including quintessence, Chaplygin gas, etc.
\citep{sahni03,alam03a}.  The ansatz (\ref{eq:taylor}) is equivalent
to the following expansion for DE
\beq\label{eq:DE1}
\rho_{\rm DE} = \rhoc (A_0 + A_1x + A_2 x^2 + A_3x^3) \,\,,
\eeq
where $\rhoc=3 H_0^2/(8 \pi G)$ is the present day critical density.
The condition $A_3 \geq 0$ allows $\rho_{\rm DE}$ to {\em mimic} the
properties of dark matter at large redshifts ($A_3 \ll 1$ follows from
large scale structure constraints).  From (\ref{eq:taylor}) and
(\ref{eq:DE1}) we find $\om = \Omega_{0{\rm m}} + A_3$, \ie the value
of $\om$ in (\ref{eq:taylor}) can be slightly larger than
$\Omega_{0{\rm m}}$ in this case.

Substituting (\ref{eq:taylor}) into the expression for the
luminosity distance we get
\beq
\frac{d_L(z)}{1+z} =  \frac{c}{H_0}\int_1^{1+z} \frac{dx}{\sqrt{\om x^3 +
A_0 + A_1 x + A_2 x^2}}\,\,.
\label{eq:lumdis}
\eeq
The parameters $A_0, A_1, A_2$ are determined by fitting
(\ref{eq:lumdis}) to supernova observations using a maximum likelihood
technique. This ansatz has only three free parameters $(\om, A_1,
A_2)$ since $A_0 + A_1 + A_2 = 1-\om$ for a flat universe.  A note of
caution: since the ansatz (\ref{eq:DE1}) is a truncated Taylor
expansion in $x=1+z$ its range of validity is $z \lleq {\rm few}$,
consequently the ansatz-derived $H(z)$ and $d_L(z)$ should not be used
at higher redshifts.

Note that the weak energy condition for dark energy $\rho_{\rm DE}\ge
0,~\rho_{\rm DE}+p_{DE}\ge 0$ has the following form for the ansatz
(\ref{eq:taylor}) :
\beq
\label{eq:WEC}
A_0 + A_1 x +  A_2 x^2 \ge 0,~~A_1 + 2A_2 x \ge 0~,
\eeq
provided we assume that the $\om x^3$ term in (\ref{eq:taylor})
is totally due to non-relativistic dark matter and does not include
any contribution from dark energy.  The demand that the WEC
(\ref{eq:WEC}) be satisfied for all $x\ge 0$ (i.e. in the past as well
as in the future) requires $A_0,A_1,A_2$ to be non-negative.  However,
the demand that the WEC (\ref{eq:WEC}) be satisfied in the past ($x
\geq 1$) but not necessarily in the future, leads to the somewhat
weaker constraint
\beq\label{eq:WEC1}
A_1+2A_2 \ge 0,~~ A_2 \geq 0~~.
\eeq
(Models in which $\rho_{\rm DE}(z) < 0$ for $z < 0$ and which violate
the WEC in the future, have been discussed in
\cite{felder02,kallosh02,alam03}.)

\begin{figure}
\begin{center}
\vspace{-0.05in}
\epsfxsize=2.4in
\epsffile{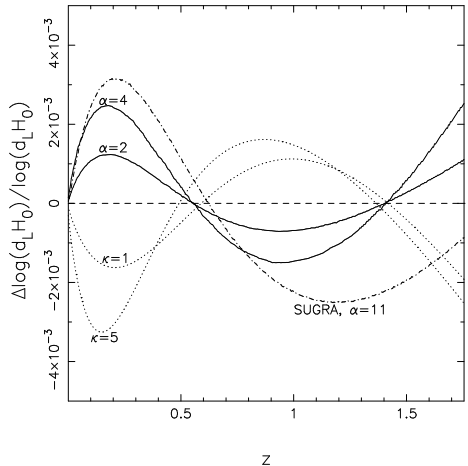}
\end{center}
\caption{\small
The fractional deviation $\Delta{\rm log}(d_L H_0)/{\rm log}(d_L H_0)$
between actual value and that calculated using the
ansatz~(\ref{eq:taylor}) over redshift for different models of dark
energy with $\omt=0.3$. The solid lines represent quintessence tracker
models for potential $V=V_0/\phi^{\alpha}$, with $\alpha=2$ and $4$.
The dotted lines show the deviation for Chaplygin Gas models with
$\kappa=1$ and $5$ (where $\kappa$ is the ratio between CDM and
Chaplygin gas densities at the commencement of the matter dominated
epoch).  The dot-dashed line represents the SUGRA potential,
$V=\left(M^{4+\alpha}/\phi^{\alpha}\right) {\rm
  exp}[\frac{1}{2}\left(\phi/M_{Pl}\right)^2]$, with $M=1.6\times
10^{-8} M_{Pl}, \alpha=11$. The dashed horizontal line represents zero
deviation from model values, which is true for $\l$CDM, and $w=-1/3,
w=-2/3$ quiessence models.  }
\label{fig:err}
\end{figure}

The presence of the term $\om x^3$ in (\ref{eq:taylor}) has two
important consequences: (i) It ensures that the the universe transits
to a matter dominated regime at early times ($z \gg 1$), (ii) It
allows us to incorporate information (available from other data sets)
regarding the current value of the matter density in the universe.
This information can be used to perform a maximum likelihood analysis
with the introduction of suitable priors on $\omt$. In further
analysis we will assume that the $\om x^3$ term in (\ref{eq:taylor})
does not include any contribution from dark energy.

We have also studied simple extensions of the ansatz (\ref{eq:taylor})
by adding new terms $A_{-1}x^{-1}$ and $A_4x^4$.  The $A_{-1}x^{-1}$
term allows $w(z)$ to become substantially less than $-1$, thereby
providing greater leeway to phantom models. The $A_4x^4$ term allows
DE to evolve towards equations of state which are more stiff than dust
($w = 0$); its role is therefore complementary to that of
$A_{-1}x^{-1}$.  Despite the inclusion of these new terms, our best
fit to the supernova data presented below does not change
significantly(choosing $A_{-1} = 0.0003$ and $A_4=0.008$), which
points to the robustness of the ansatz (\ref{eq:taylor}) for the given
data set.

We should add that our reason for choosing an ansatz to fit $H(z)$
rather than some other cosmological quantity was motivated by the fact
that the Hubble parameter is directly related to a fundamental
physical quantity -- the Ricci tensor, and is therefore likely to
remain meaningful even when other quantities (such as the equation of
state) become `effective'. (This happens for instance, in the case of
the braneworld models of dark energy discussed in \cite{DDG,ss02a}.)

The rationale for choosing a three parameter ansatz for $H(z)$ is the
following.  The observed luminosity distance determined using type Ia
supernovae is rather noisy, therefore in order to determine the Hubble
parameter from $D_L(z)$ and following that the equation of state, one
must take two derivatives of a noisy quantity. This difficulty can be
tackled in two possible ways: (i) either one smoothes the data over
some interval $\Delta t$ (binning is one possibility), or (ii) we may
choose to smooth `implicitly' by parameterizing $H(z)$ through an
appropriate fitting function.  The number of free parameters ${\cal
  N}$ in the fit to $H(z)$ will be related to the smoothing interval
$\Delta z$ through $\Delta z = z_{\rm max}/{\cal N}$. Increasing
${\cal N}$ implies decreasing $\Delta z$ which results in a rapid
growth of errors through $\Delta H(z) \propto (\Delta z)^{-3/2}$, and
$\Delta w(z) \propto (\Delta z)^{-5/2}$ \citep{teg02}, therefore in
order not to loose too much accuracy in our reconstruction we
considered 3 parameter fits for $H(z)$ in our paper (these correspond
to 2 parameter fits for $w(z)$).

We now test the usefulness of the ansatz~(\ref{eq:taylor}) in
reconstructing different dark energy models. The ansatz returns exact
values for $\l$CDM, and $w=-1/3$, $w=-2/3$ quiessence models.  In
figure~\ref{fig:err} we show the accuracy of the ansatz
(\ref{eq:taylor}) when applied to several other dark energy models
such as tracker quintessence, the Chaplygin gas and super-gravity
(SUGRA) models. We plot the deviation of ${\rm log}(d_L H_0)$ (which
is the measured quantity for SNe) obtained with the
ansatz~(\ref{eq:taylor}) from the actual model values. Clearly the
ansatz performs very well over a significant redshift range for
$\omt=0.3$ (Also see appendix~\ref{sec:fits}). In fact, in the
redshift range where SNe data is available, the ansatz recovers these
models of dark energy with less than $0.5\%$ errors. However it would
be appropriate to add a note of caution at this point.  Although
figure~\ref{fig:err} clearly demonstrates the usefulness of the ansatz
for some DE models, its performance vis-a-vis other models of DE is by
no means guaranteed. By its very construction the ansatz
(\ref{eq:taylor}) is expected to have limitations when describing
models with a fast phase transition \citep{bassett} as well as rapidly
oscillating quintessence models \citep{sahwang}.  (The ansatz
(\ref{eq:taylor}) can give reasonable results even for these models
provided the resulting DE behaviour is suitably smoothed.)  For this
reason, although the bulk of our analysis will be carried out using
(\ref{eq:taylor}), we shall supplement it when necessary with other
fitting functions, which will provide us with an independent means
with which to test the robustness of our reconstruction exercise.

{\bf Methodology :}

For our primary reconstruction, we use a subset of 172 type Ia
Supernovae, obtained by imposing constraints $A_V < 0.5$ and $z >
0.01$ on the 230 SNe sample, as in the primary fit of \cite{tonry03}.
For the ansatz (\ref{eq:lumdis}), we require to fit four parameters:
($H_0, \omt, A_1, A_2$). We may use prior information on $H_0$ ($H_0 =
72 \pm 8 \ {\rm km \ s^{-1} \ Mpc^{-1}}$, \cite{HST}) and $\omt$ ($\omt
h = 0.2 \pm 0.03$, \cite{2dF}). \footnote{One should note however that
  the prior on $\omt$ is not model independent since it relies on the
  $\l$CDM model to project from redshift space to real space. Results
  coming from the use of this prior should therefore not be taken too
  literally in the present context. See \cite{kunz03} for an
  interesting discussion of related issues.}

The measured quantity for this data is $y=<{\rm log}(d_L H_0)>$,
therefore the likelihood function is given by
\ber
{\cal L} &=& {\cal N} \exp \left ( - \frac{\chi^2}{2} \right ) \,,\\
\chi^2 &=& \sum_{i=1}^{172} \left( \frac{y_i-y_{\rm fit}(H_0, \omt, A_1, A_2)}{\sigma_i} \right)^2 \,,
\eer
where ${\cal N}$ is a normalisation constant. Therefore, the
probability distribution function in the four-space $(H_0, \omt,
A_1, A_2)$ is
\beq
{\rm P}(H_0, \omt, A_1, A_2) \propto \exp \left (
-\frac{\chi^2}{2} \right ) {\rm Pr}(\omt h) {\rm Pr}(H_0) \,.
\eeq
where ${\rm Pr}$ refers to the priors applied on the parameters of the
system.

Our goal is to reconstruct cosmological parameters such as the
equation of state $w(z) = w(z;\omt,A_1,A_2)$, therefore we marginalise
over $H_0$ and obtain the probability distribution function in the
$(\omt, A_1, A_2)$ space:
\beq
\tilde{P}(\omt, A_1, A_2) = \int {\rm P}(H_0, \omt, A_1, A_2) dH_0 \,.
\eeq
In order to do this, we have to define the bounds of a
four-dimensional volume in $(H_0, \omt, A_1, A_2)$. The bounds of $H_0$
are taken at $5 \sigma$ of the HST prior. For $\omt$, the natural
choice is $0 \leq \omt \leq 1$. It is not immediately obvious what the
bounds should be for $A_1, A_2$. We choose a sufficiently large
rectangular grid for $A_1,A_2$ (roughly corresponding to $-6 \lleq w_0
\lleq 5$) which includes most known models of dark energy. This bound
is merely a matter of convenience and does not affect our results in
any way. After marginalisation, we have a grid in $(\omt, A_1, A_2)$
space on which $\tilde{P}(\omt, A_1, A_2)$ is specified at each point.
We may now proceed in two ways. Firstly, we may choose to fix $\omt$ at
a suitable constant value (\eg $\omt=0.3$) thereby obtaining a grid in
the $(A_1, A_2)$ plane with $P$ (the probability if $\omt$ is known to
be an exact value) defined at each point. For a particular redshift,
we may then calculate $w(z;\omt, A_1, A_2)$ at each point of the grid.
This would yield results that would hold true if $\omt$ were known
exactly.  Instead of using the exact value of $\omt$, we may use the
prior information about it available to us ($\omt h = 0.2 \pm 0.03$),
and calculate $w(z;\omt, A_1, A_2)$ at each point of a
three-dimensional grid, the probability $\tilde{P}$ at each point
being known.  Therefore, at any given redshift $z$, $w(\omt, A_1, A_2)$
can be tagged with a numerical value $ \tilde{P}(\omt, A_1, A_2)$.
Starting from the best-fit $w(z)$ (the value at the peak of the
probability distribution), we may move down on either side till $34
\%$ of the total area is enclosed under the curve, thus obtaining
asymmetric $1 \sigma$ bounds on $w(z)$. The $2 \sigma, \ 3 \sigma$
bounds can be similarly obtained.

\begin{figure}
\centering
\begin{center}
\vspace{-0.05in}
\epsfxsize=2.4in
\epsffile{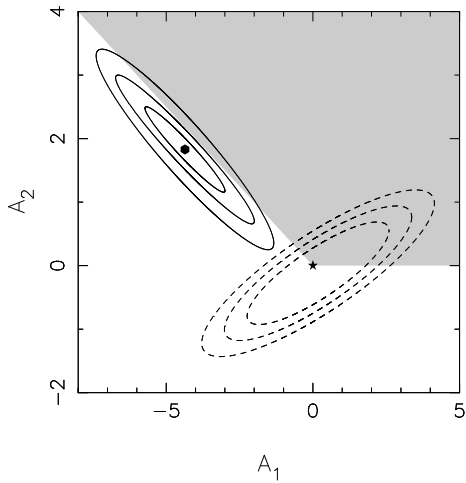}
\end{center}
\caption{\small
The ($A_1,A_2$) parameter space for the ansatz~(\ref{eq:taylor}).
The light grey shaded area shows the allowed region if dark energy
satisfies the weak energy condition both currently and in the past:
$w(z) \geq -1, z \geq 0$.  The $\chi^2$ surface has two minima, a
shallow minimum at $A_1=0.177, A_2=-0.119$ with $\chi^2_{\rm shallow}
= 1.0402$ and a deeper minimum at $A_1=-4.360, A_2=1.829$ with
$\chi^2_{\rm deep} = 1.0056$. The deeper minimum is marked by a
bullet.  The solid contours surrounding the deeper minimum are
$1\sigma, 2\sigma, 3\sigma$ contours of constant $\Delta\chi^2$ where
$\Delta\chi^2 = \chi^2 - \chi^2_{\rm deep}$.  Similarly the dashed
contours surrounding the shallower minimum are $1\sigma, 2\sigma,
3\sigma$ contours of constant $\Delta\chi^2$ where $\Delta\chi^2 =
\chi^2 - \chi^2_{\rm shallow}$.  Note that the $\l$CDM model
(marked by a solid star) corresponds to $A_1 = A_2 = 0$ which is very
close to the shallow minimum.}
\label{fig:chi}
\end{figure}

\begin{figure}
\centering
\begin{center}
\vspace{-0.05in}
\epsfxsize=2.4in
\epsffile{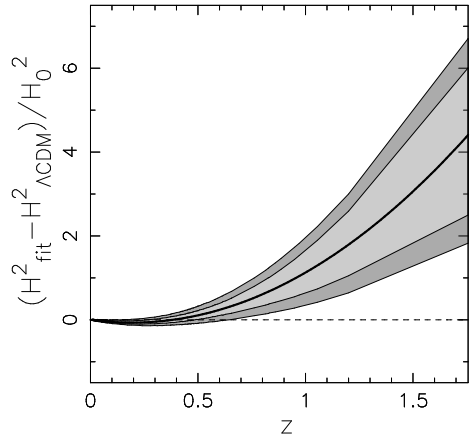}
\end{center}
\caption{\small
The deviation of $H^2/H_0^2$ from corresponding $\l$CDM values over
redshift for the ansatz~(\ref{eq:taylor}). The thick solid line shows
the best-fit, the light grey contour represents the $1\sigma$
confidence level, and the dark grey contour represents the $2\sigma$
confidence level around the best-fit.  The dashed horizontal line
denotes $\l$CDM. $\omt = 0.3$ is assumed.}
\label{fig:dev}
\end{figure}

{\bf Results :}

We first show preliminary results for which the matter density is
fixed at a constant value of $\omt=0.3$. A detailed look at the
$\chi^2$ surface in the ($A_1, A_2$) plane (figure~\ref{fig:chi})
reveals the existence of two minima in $\chi^2$, a shallower one close
to $\l$CDM ($A_1=0.177, A_2=-0.119, w_0=-1.03 , \chi^2=1.0402$), and a
deeper minimum at $A_1=-4.360, A_2=1.829, w_0=-1.33, \chi^2=1.0056$.
We would like to draw the readers attention to the fact that imposing
the prior $w(z) \geq -1 \ (z \geq 0)$ amounts to disallowing a
significant region of parameter space (the unshaded region in
figure~\ref{fig:chi}).  Consequently an analysis which assumes $w(z)
\geq -1$ loses all information about the region $2\sigma$ around the
deeper minimum ! Since we have no reason (observational or
theoretical) to favour either minimum over the other, we shall always
choose the deeper minimum as our best-fit in all the subsequent
calculations.

\begin{figure}
\centering
\begin{center}
\vspace{-0.05in}
\epsfxsize=2.4in
\epsffile{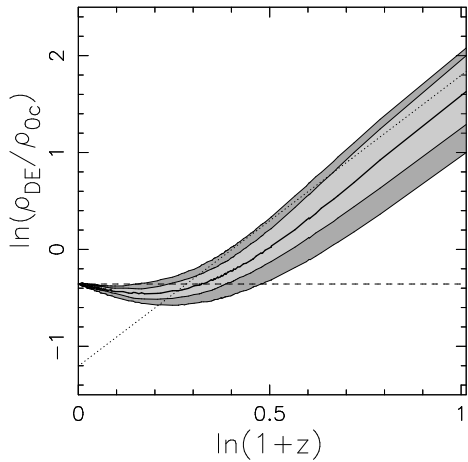}
\end{center}
\caption{\small
The logarithmic variation of dark energy density $\rho_{\rm DE}/\rhoc$
(where $\rhoc=3 H_0^2/8 \pi G$ is the present day critical density)
with redshift for the ansatz~(\ref{eq:taylor}). The thick solid line
shows the best-fit, the light grey contour represents the $1\sigma$
confidence level, and the dark grey contour represents the $2\sigma$
confidence level around the best-fit.  The dashed horizontal line
denotes $\l$CDM and the dotted line represents matter density $\omt
(1+z)^3$, $\omt = 0.3$ is assumed.}
\label{fig:dens}
\end{figure}

In the figure~\ref{fig:dev}, we plot the deviation of the squared
Hubble parameter $H^2/H_0^2$ from $\l$CDM over redshift for the
best-fit. We note that the quantity $H^2/H_0^2$ has a simple linear
relationship with the parameters of the fit (Eq~\ref{eq:taylor}),
therefore the errors in this quantity increase with redshift. Another
quantity of interest is the energy density of dark energy. For this
ansatz, $\tilde \rho_{\rm DE}=\rho_{\rm DE}/\rhoc=A_0+A_1 x+A_2 x^2$
(where $\rhoc=3 H_0^2/8 \pi G$ is the present day critical density).
The figure~\ref{fig:dens} shows the logarithmic variation of $\tilde
\rho_{\rm DE}$ with redshift. In this figure too the errors increase
with redshift. An interesting point to note is that initially, dark
energy density decreases with redshift, showing the phantom-like
nature ($w < -1$) of dark energy at lower redshifts of $z \lleq 0.25$,
while at higher redshifts, the dark energy density begins to track the
matter density. Before moving on to the second derivative of the
luminosity distance (\eg the equation of state) we may obtain more
information from the dark energy density by considering a weighted
average of the equation of state :
\beq
1+\bar{w}=\frac{1}{\Delta \ {\rm ln}(1+z)} \int (1+w(z)) \frac{dz}{1+z} \,\,,
\label{eq:w_avg1}
\eeq
where $\Delta$ denotes the total change of the variable between
integration limits. This quantity can be elegantly expressed in terms
of the difference in energy densities over a range of redshift as
\beq
1+\bar{w}=\frac{1}{3} \frac{\Delta \ {\rm ln}\tilde \rho_{\rm DE}}{\Delta \ {\rm ln}(1+z)}\,\,. 
\label{eq:w_avg2}
\eeq
Thus the variation in the dark energy density depicted in
figure~\ref{fig:dens} is very simply related to the weighted average
equation of state !

\begin{table}
\begin{center}
\caption{
The weighted average $\bar{w}$ (eq~\ref{eq:w_avg1}) over specified
redshift ranges. The best-fit value and $1\sigma$ and $2\sigma$
deviations from the best-fit are shown.}
\label{tab:w_avg}
\begin{tabular}{cccc}
$\Delta z$&$\bar{w}$&$1\sigma$&$2\sigma$\\
\hline
$0-0.414$&$-0.969$&${ }^{+0.120}_{-0.089}$&${ }^{+0.198}_{-0.199} $\\
\vspace{0.01in}\\
$0.414-1$&$-0.108$&${ }^{+0.230}_{-0.240}$&${ }^{+0.242}_{-0.360}$\\
\vspace{0.01in}\\
$1-1.756$&$0.069$&${ }^{+0.100}_{-0.080}$&${ }^{+0.130}_{-0.180}$\\
\end{tabular}
\end{center}
\end{table}

\begin{figure*}
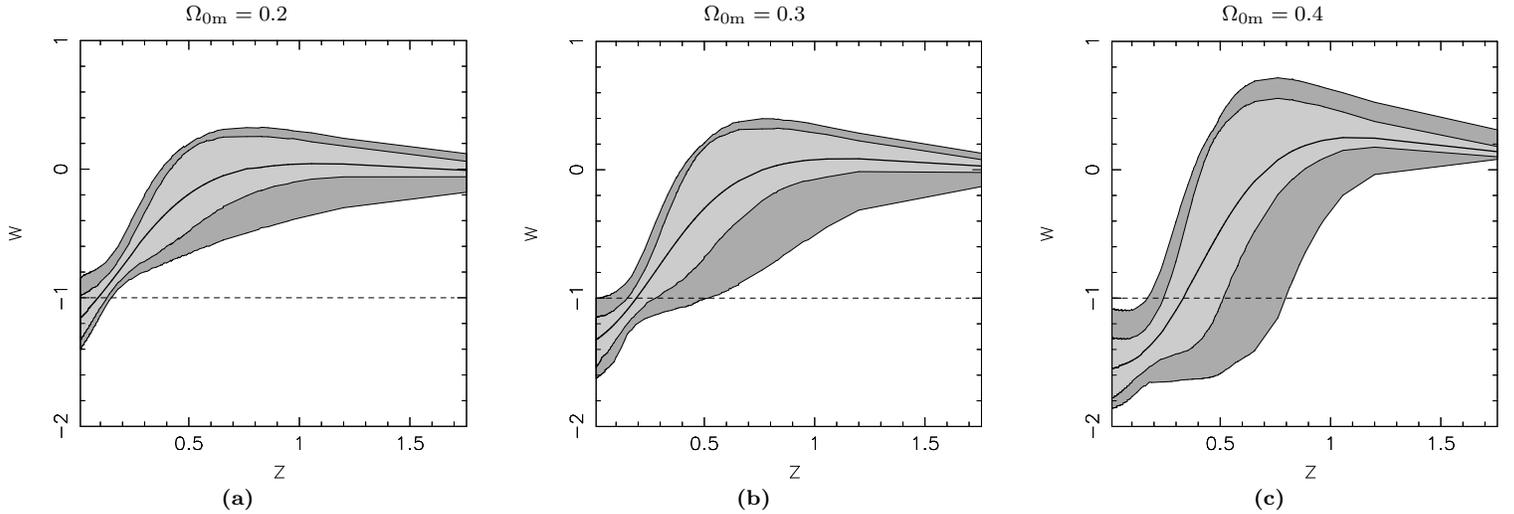

\centering
\begin{center}
\vspace{-0.05in}
\centerline{\mbox{\hspace{0.in} $\omt=0.2$ \hspace{2.1in} $\omt=0.3$ \hspace{2.1in} $\omt=0.4$}}
$\begin{array}{@{\hspace{-0.4in}}c@{\hspace{0.3in}}c@{\hspace{0.3in}}c}
\multicolumn{1}{l}{\mbox{}} &
\multicolumn{1}{l}{\mbox{}} &
\multicolumn{1}{l}{\mbox{}} \\ [-0.5cm]
\epsfxsize=2.4in
\epsffile{h_exp_w_mi_tot.epsi} &
\epsfxsize=2.4in
\epsffile{h_exp_w_tot.epsi} &
\epsfxsize=2.4in
\epsffile{h_exp_w_pl_tot.epsi} \\
\mbox{\bf (a)} & \mbox{\bf (b)} & \mbox{\bf (c)}
\end{array}$
\end{center}
\caption{\small
The evolution of $w(z)$ with redshift for different values of $\omt$.
The reconstruction is done using the polynomial fit to dark energy,
equation~(\ref{eq:taylor}). In each panel, the thick solid line shows
the best-fit, the light grey contour represents the $1\sigma$
confidence level, and the dark grey contour represents the $2\sigma$
confidence level around the best-fit. The dashed line represents
$\l$CDM. No priors are assumed on $w(z)$. The $\chi^2$ per degree
of freedom for each case is given in Table~\ref{tab:chi}. }
\label{fig:total}
\end{figure*}

\begin{table}
\begin{center}
\caption{
$\chi^2$ per degree of freedom for best-fit and $\l$CDM models.
$w_0$ is the present value of the equation of state of dark energy in
best-fit models. $\chi^2_{\rm Pr}$ refers to the best fit after
imposition of the WEC prior $w(z) \geq -1 \ (z \geq 0)$.}
\label{tab:chi}
\begin{tabular}{@{\hspace{-0.015in}}ccccc@{\hspace{-0.045in}}cc}
$ $&\multicolumn{2}{c}{Best-fit}& \multicolumn{2}{c}{Confidence levels}&$w(z) \geq -1$&$\l$CDM \\
$\omt$&$ w_0$&$\chi^2_{\rm min}$&$\chi^2_{1\sigma}$&$\chi^2_{2\sigma}$&$\chi^2_{\rm Pr}$&$ \chi^2$\\
\hline
$0.10$&$-1.093$&$1.0077$&$1.0213$&$1.0442$&$1.0359$&$1.1242$\\
$0.20$&$-1.198$&$1.0071$&$1.0207$&$1.0436$&$1.0384$&$1.0663$\\
$0.30$&$-1.334$&$1.0056$&$1.0192$&$1.0421$&$1.0409$&$1.0417$\\
$0.40$&$-1.470$&$1.0043$&$1.0179$&$1.0408$&$1.0578$&$1.0638$\\
$0.50$&$-1.606$&$1.0038$&$1.0174$&$1.0403$&$1.0912$&$1.1168$\\
\end{tabular}
\end{center}
\end{table}

In table~\ref{tab:w_avg} we show the values of $\bar{w}$ obtained
using different ranges in redshift for our best-fit with corresponding
$1\sigma$ and $2\sigma$ errors. We have taken the ranges of
integration to be approximately equally spaced in ln$(1+z)$, with the
upper limit set by the furthest supernova known at present. The values
of $\bar{w}$ may be calculated using the equation~(\ref{eq:w_avg1})
(which uses the second derivative of the luminosity distance), or they
can simply be read off from figure~\ref{fig:dens} using
equation~(\ref{eq:w_avg2}). From this table, a ``metamorphosis'' in
the properties of dark energy occurring somewhere between $z\sim0$ and
$z\sim1$ can be clearly seen (note that, effectively, one needs to
differentiate $d_L(z)$ only once to come to this conclusion).

\begin{figure*}
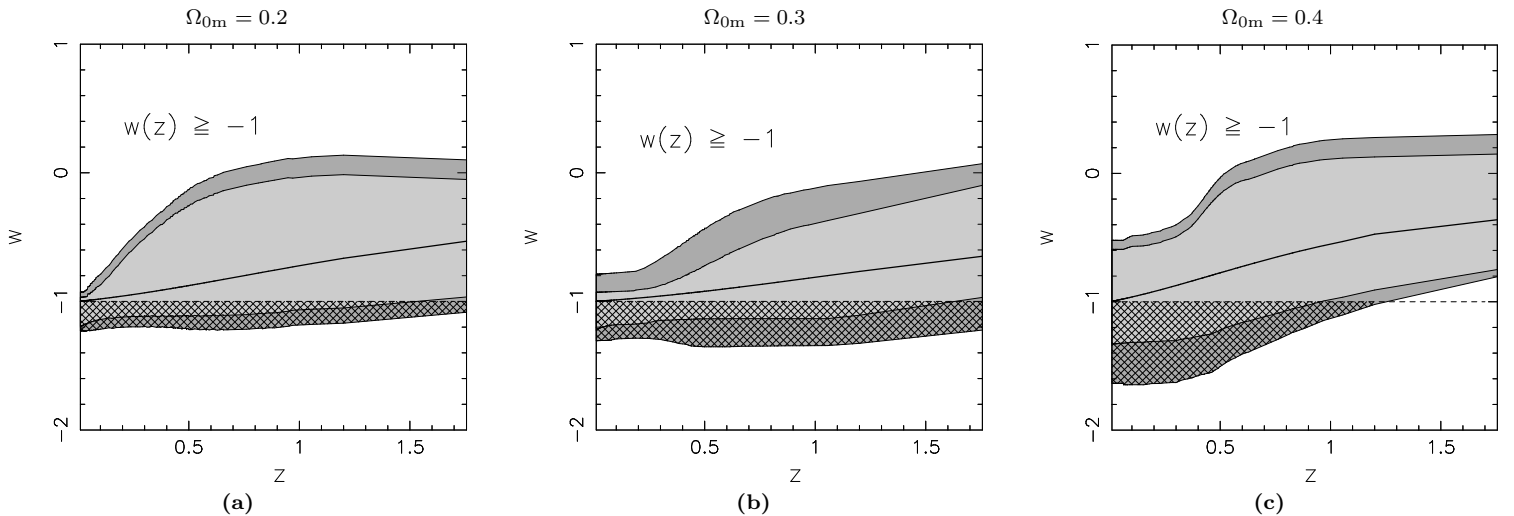

  \centering
\begin{center}
\vspace{-0.05in}
\centerline{\mbox{\hspace{0.in} $\omt=0.2$ \hspace{2.1in} $\omt=0.3$ \hspace{2.1in} $\omt=0.4$}}
$\begin{array}{@{\hspace{-0.4in}}c@{\hspace{0.3in}}c@{\hspace{0.3in}}c}
\multicolumn{1}{l}{\mbox{}} &
\multicolumn{1}{l}{\mbox{}} &
\multicolumn{1}{l}{\mbox{}} \\ [-0.5cm]
\epsfxsize=2.4in
\epsffile{h_exp_w_mi_cons.epsi} &
\epsfxsize=2.4in
\epsffile{h_exp_w_cons.epsi} &
\epsfxsize=2.4in
\epsffile{h_exp_w_pl_cons.epsi} \\
\mbox{\bf (a)} & \mbox{\bf (b)} & \mbox{\bf (c)}
\end{array}$
\end{center}
\caption{\small
The evolution of $w(z)$ with redshift for different values of $\omt$,
using the prior $w(z) \geq -1, z \geq 0$. The reconstruction is done
using the polynomial fit to dark energy, equation~(\ref{eq:taylor}). In
each panel, the thick solid line shows the best-fit, the light grey
contour represents the $1\sigma$ confidence level, and the dark grey
contour represents the $2\sigma$ confidence level around the best-fit.
The hatched region is forbidden by the prior $w(z) \geq -1$. The
dashed line represents $\l$CDM.  The $\chi^2$ per degree of
freedom for each case is given in Table~\ref{tab:chi}. }
\label{fig:constr}
\end{figure*}

\begin{figure*}
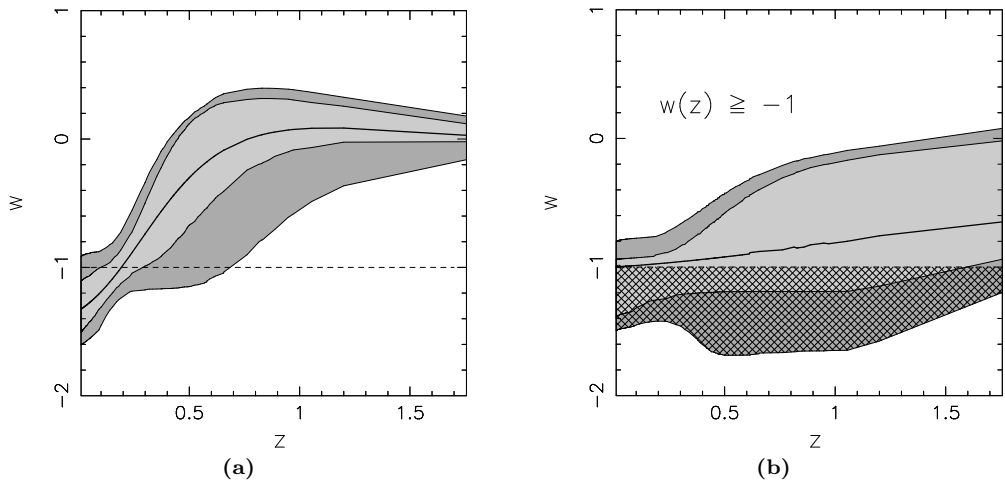

\centering
\begin{center}
\vspace{-0.05in}
$\begin{array}{c@{\hspace{0.4in}}c}
\multicolumn{1}{l}{\mbox{}} &
\multicolumn{1}{l}{\mbox{}} \\ [0.0cm]
\epsfxsize=2.4in
\epsffile{h_exp_w_m.epsi} &
\epsfxsize=2.4in
\epsffile{h_exp_w_m_cons.epsi} \\
\mbox{\bf (a)} & \mbox{\bf (b)}
\end{array}$
\end{center}
\caption{\small
Evolution of $w(z)$ with redshift using the 2dF prior $\omt h=0.2 \pm
0.03$, with (a) no priors on $w(z)$, and (b) the prior $w(z) \geq -1,
~z \geq 0$. The reconstruction is done using the polynomial fit to
dark energy, equation~(\ref{eq:taylor}). In both panels, the thick solid
line represents the best-fit, the dashed line represents $\l$CDM,
the light grey contour represents $1\sigma$ confidence level, and the
dark grey contour represents the $2\sigma$ confidence level. In the
right hand panel, the hatched region is forbidden by the prior $w(z)
\geq -1$.}
\label{fig:margin}
\end{figure*}

We now reconstruct the equation of state of dark energy which, for the
ansatz~(\ref{eq:taylor}), has the form
\beq\label{eq:state2} 
w(z) = -1 + \frac{x}{3}\frac{A_1 + 2A_2 x}{A_0 + A_1 x + A_2 x^2}, ~~ x = 1+z~.
\eeq 
(Note that, since $w(z)$ was derived from the ansatz
(\ref{eq:taylor}), its domain of validity is $z \lleq {\rm few}$.)  In
figures~\ref{fig:total} (a), (b), (c), we show the evolution of the
equation of state $w(z)$ with redshift for different values of $\omt$.
The $1\sigma$ and $2\sigma$ limits are shown in each case.  The
$\chi^2$ per degree of freedom for the best-fit for the different
cases is given in Table~\ref{tab:chi}. We find that for $0.2 < \omt <
0.4$, the behaviour of the equation of state does not change
significantly with change in the matter density.  However, for larger
values of $\omt$, a \emph{smaller} current value of $w_0=w(z=0)$ is
preferred.  In all three cases considered, the present value of the
equation of state is $w_0 \lleq -1.2$ for the best-fit, and the
equation of state rises steeply from $w \lleq -1.2$ to $w \simeq 0$
with redshift. In fact, the behaviour of $w$ appears to be extremely
different from that in $\l$CDM ($w=-1$). We note here that, for this
analysis, the errors on $w$ appear to decrease with redshift. This may
appear counter-intuitive, since there are fewer SNe at higher
redshifts, but this is merely a construct of the fact that $w$ depends
non-linearly on the parameters of the ansatz (see
appendix~\ref{sec:errors}).

Quintessence models satisfy the weak energy condition (WEC) $\rho + p
= {\dot\phi}^2 \geq 0$ and it would be interesting to see how the
imposition of the WEC as a prior on the equation of state will affect
the results of our analysis. We therefore perform the same analysis as
above with the added constraint $w_0 \geq -1$ (note that this implies
$w(z)\ge -1$ for all $z \geq 0$ for our fitting function of $H(z)$
provided $A_2 \geq 0$).  The results are shown in the
figures~\ref{fig:constr} (a), (b), (c).  We see that in this case the
errors are larger and the evolution of the equation of state with
redshift follows a much gentler slope. Such an equation of state would
be largely consistent with the cosmological constant model. (These
results are in broad agreement with an earlier analysis of
\cite{saini00} in which a smaller SNe data set was used and a different
ansatz for the luminosity distance was applied.)

In Table~\ref{tab:chi}, we show how the $\chi^2$ for the best-fit
evolving dark energy models compare with that for $\l$CDM. We
find that $\chi^2_{\l {\rm CDM}} > \chi^2_{\rm best-fit}$ always.
For $\omt=0.3$, the value of $\chi^2_{\l {\rm CDM}}$ is just
within $2\sigma$ of the best fit $\chi^2$, but for $\omt=0.2$, or
$\omt=0.4$, $\chi^2_{\l {\rm CDM}}$ is outside the $2\sigma$
limits of the best-fit. It is also noteworthy that when the prior
$w(z) \geq -1 \ (z \geq 0)$ is used, the best-fit model has a slowly
evolving equation of state with $w_0=-1$ and the $\chi^2$ for the
best-fit becomes smaller for a smaller value of the matter density.
When no priors are assumed on $w$, the trend reverses, and better fits
are obtained for larger values of $\omt$. From this it appears that at
least at $1\sigma$ the evolving dark energy model is favoured over
$\l$CDM, and it does as well, if not better at the $2\sigma$
level, depending upon the value of the present-day matter density.

{\bf Using Priors on $\omt$ : }

Instead of assuming an exact value for $\omt$, which is somewhat
optimistic given the present observational scenario, we may use the
2dF prior on $\omt$ and calculate $w(z)$ as a function of $(\omt, A_1,
A_2)$. It should be noted here that the 2dF error bars on $\omt h$ have
been calculated using 2dF data in conjunction with CMB, and this
calculation assumes a $\l$CDM model, therefore this prior should
be used more as a benchmark for the value of $\omt$ rather than as an
absolute when considering evolving dark energy models. The resultant
``marginalised'' $w$ is shown as a function of the redshift in
figure~\ref{fig:margin} (a). The nature of the equation of state for
the analysis with the added prior $w(z) \geq -1 \ (z \geq 0)$ is shown
in figure~\ref{fig:margin} (b). We find that the general nature of
evolution of the equation of state is not changed by adding this extra
information on the matter density. If no priors are assumed on the
equation of state to begin with, $w(z)$ still rises sharply from $w_0
\lleq -1$ up to $w \simeq 0$ at maximum redshift and the analysis
appears to favour a fast-evolving equation of state of dark energy
over the standard $\l$CDM model. If a prior $w \geq -1$ is
assumed, then the marginalised equation of state is more consistent
with the cosmological constant. From this we see that {\em
  marginalisation over $\omt$ does not lead to any significant change
  in our results}. In the subsequent sections, we will show our
results for $\omt=0.3$.

From the above analysis, we find, therefore, that our results change
significantly depending upon whether or not the prior $w \geq -1$ is
imposed. We saw earlier that in the absence of any prior on $w(z)$,
the best-fit equation of state rose from $w \lleq -1$ at $z=0$ to $w
\simeq 0$ at $z \sim 1$. By imposing a prior on the equation of state,
we effectively screen off a sizeable part of the parameter space (see
figure~\ref{fig:chi}), and therefore the reconstruction is forced to
choose its best-fit away from the true minima of the $\chi^2$ surface.
The effect of imposing a prior on $w(z)$ is therefore to make the
best-fit $w(z)$ grow much more slowly with $w = -1$ being preferred at
$z = 0$.  Our results show that the reconstructed equation of state
with the prior $w \geq -1$ is in good agreement with a cosmological
constant at the $68\%$ CL. However, if no prior is imposed, then the
steeply evolving dark energy models are favoured over the cosmological
constant at $1\sigma$, and are at least as likely as the cosmological
constant at the $2\sigma$ level.

\begin{figure}
\centering
\begin{center}
\vspace{-0.05in}
\epsfxsize=2.4in
\epsffile{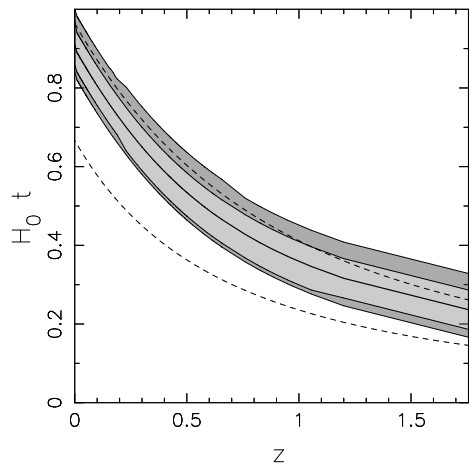}
\end{center}
\caption{\small
The age of the universe, $H_0 t(z)$, is shown as a function of the
redshift for $\omt=0.3$. The reconstruction is done using the
polynomial fit to dark energy, equation~(\ref{eq:taylor}). The thick
solid line represents the best-fit, the light grey contour represents
$1\sigma$ confidence level, and the dark grey contour represents the
$2\sigma$ confidence level.  The upper (lower) dashed line represents
$\l$CDM (SCDM).  }
\label{fig:age}
\end{figure}

{\bf Age and Deceleration Parameter of the Universe: }

We may also use this ansatz to calculate other quantities of interest,
such as the age of the universe, $t(z)$, and the deceleration
parameter, $q(z)$:
\ber
t(z) &=& \int_{1+z}^{\infty} \frac{dx}{x H(x)} \\
q(z) &=& -\frac{\ddot a}{aH^2} \equiv \frac{H^{\prime}}{H}x -1 \,\,.
\eer
where $x=1+z$.

In figure~\ref{fig:age} we plot the evolution of the age of the
universe with redshift. We find that the best-fit age of the universe
today is $t_0=12.8$ Gyrs if the Hubble parameter is taken to be
$H_0=72\ {\rm km \ s^{-1} \ Mpc^{-1}}$, which is slightly lower than
the age of a $\l$CDM universe, $t_0=13.4$ Gyrs (both values are
for $\omt = 0.3$).  At the $2\sigma$ level, the age of the universe
today would vary between $11.2 \leq t_0 \leq 13.6$ Gyrs.

\begin{figure}
\centering
\begin{center}
\vspace{-0.05in}
\epsfxsize=2.4in
\epsffile{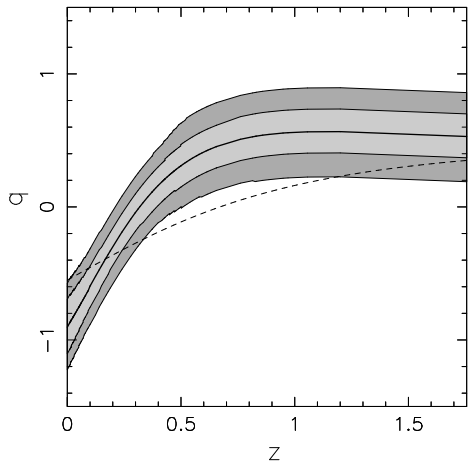}
\end{center}
\caption{\small
Evolution of deceleration parameter of the universe, $q(z)$ with
redshift for $\omt=0.3$. The reconstruction is done using the
polynomial fit to dark energy, equation~(\ref{eq:taylor}). The thick
solid line represents the best-fit, the light grey contour represents
$1\sigma$ confidence level, and the dark grey contour represents the
$2\sigma$ confidence level. The dashed line represents $\l$CDM.
}
\label{fig:dec}
\end{figure}

Figure~\ref{fig:dec} shows the evolution of the deceleration parameter
with redshift.  We find that the behaviour of the deceleration
parameter for the best-fit universe is quite different from that in
$\l$CDM cosmology.  Thus, the current value of $q_0 \simeq -0.9$ is
significantly lower than $q_0 \simeq -0.55$ for $\l$CDM (assuming $\omt
= 0.3$).  Furthermore the rise of $q(z)$ with redshift is much steeper
in the case of the best-fit model, with the result that the universe
begins to accelerate at a comparatively lower redshift $z \simeq 0.3$
(compared with $z \simeq 0.7$ for $\l$CDM) and the matter dominated
regime ($q \simeq 1/2$) is reached by $z \sim 1$.

\begin{figure*}
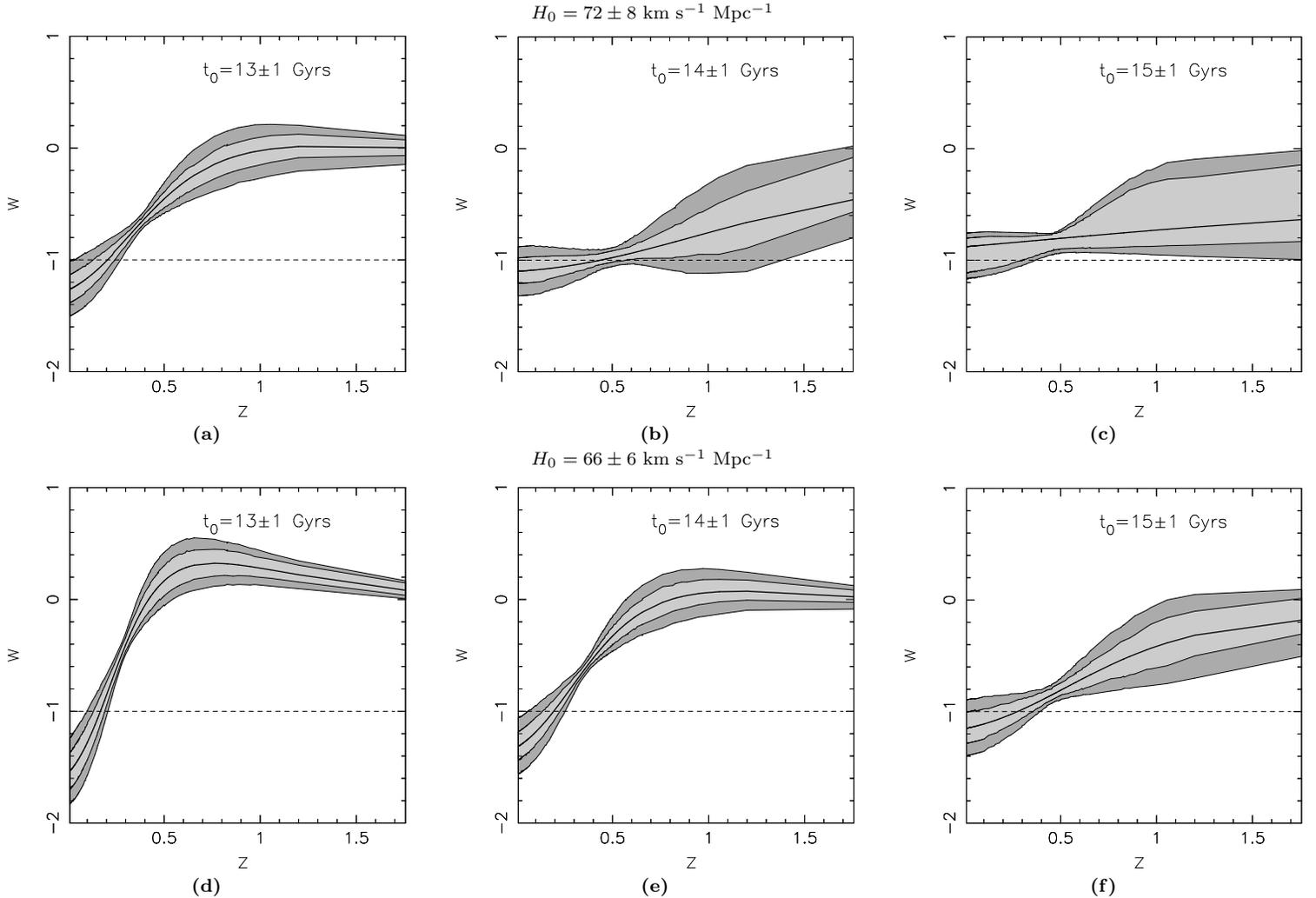

\centering
\begin{center}
\vspace{-0.05in}
\centerline{\mbox{ $H_0=72 \pm 8\ {\rm km \ s^{-1} \ Mpc^{-1}}$ }}
$\begin{array}{@{\hspace{-0.4in}}c@{\hspace{0.3in}}c@{\hspace{0.3in}}c}
\multicolumn{1}{l}{\mbox{}} &
\multicolumn{1}{l}{\mbox{}} &
\multicolumn{1}{l}{\mbox{}} \\ [-0.5cm]
\epsfxsize=2.4in
\epsffile{h_exp_w_age1.epsi} &
\epsfxsize=2.4in
\epsffile{h_exp_w_age.epsi} &
\epsfxsize=2.4in
\epsffile{h_exp_w_age2.epsi} \\
\mbox{\bf (a)} & \mbox{\bf (b)} & \mbox{\bf (c)}
\end{array}$
\vspace{0.15in}
\centerline{\mbox{ $H_0=66 \pm 6\ {\rm km \ s^{-1} \ Mpc^{-1}}$ }}
$\begin{array}{@{\hspace{-0.4in}}c@{\hspace{0.3in}}c@{\hspace{0.3in}}c}
\multicolumn{1}{l}{\mbox{}} &
\multicolumn{1}{l}{\mbox{}} &
\multicolumn{1}{l}{\mbox{}} \\ [-0.5cm]
\epsfxsize=2.4in
\epsffile{h_exp_w_age4.epsi} &
\epsfxsize=2.4in
\epsffile{h_exp_w_age3.epsi} &
\epsfxsize=2.4in
\epsffile{h_exp_w_age5.epsi} \\
\mbox{\bf (d)} & \mbox{\bf (e)} & \mbox{\bf (f)}
\end{array}$
\end{center}
\caption{\small
The evolution of $w(z)$ with redshift for $\omt=0.3$ using a Gaussian
prior on the age of the universe today: $t_0=13 \pm 1$Gyrs (figures
(a) and (d)), $t_0=14 \pm 1$Gyrs (figures (b) and (e)), and $t_0=15
\pm 1$Gyrs (figures (c) and (f)). The reconstruction is done using the
polynomial fit to dark energy, equation~(\ref{eq:taylor}). The top panel
is obtained using a Gaussian prior $H_0 =72 \pm 8\ {\rm km \ s^{-1} \
  Mpc^{-1}}$, while for the bottom panel, $H_0 =66 \pm 6\ {\rm km \
  s^{-1} \ Mpc^{-1}}$. No priors are assumed on $w(z)$. In each panel,
the thick solid line shows the best-fit, the light grey contour
represents the $1\sigma$ confidence level, and the dark grey contour
represents the $2\sigma$ confidence level around the best-fit.  The
dashed line represents $\l$CDM. The $\chi^2$ per degree of
freedom for each case is given in Table~\ref{tab:chi_pr}.}
\label{fig:age_pr}
\end{figure*}

\begin{table}
\begin{center}
\caption{
$\chi^2$ per degree of freedom for the ansatz (\ref{eq:taylor})
which best fits the SNe data after different age priors are imposed.
$w_0$ is the present value of the equation of state of dark energy in
best-fit models.}
\label{tab:chi_pr}
\begin{tabular}{cccc}
$H_0 $&$t_0$&\multicolumn{2}{c}{Best-fit}\\
${\rm km \ s^{-1} \ Mpc^{-1}}$&${\rm Gyrs}$&$ w_0$&$\chi^2_{\rm min}$\\
\hline
$72 \pm 8$&$13 \pm 1$&$-1.271$&$1.0062$\\
$$&$14 \pm 1$&$-1.099$&$1.0197$\\
$$&$15 \pm 1$&$-0.904$&$1.0407$\\
\hline
$66 \pm 6$&$13 \pm 1$&$-1.553$&$1.0139$\\
$$&$14 \pm 1$&$-1.324$&$1.0057$\\
$$&$15 \pm 1$&$-1.153$&$1.0146$\\
\end{tabular}
\end{center}
\end{table}

{\bf Using Priors on Age of the Universe : }

Important consistency checks on our best-fit Universe may be provided
by observations of the age of the Universe. Unfortunately, estimates
of the age of the universe from different methods can produce widely
varying results one reason for which is that estimates of the Hubble
parameter itself can vary significantly. For instance, the HST key
project yields $H_0=72 \pm 8\ {\rm km \ s^{-1} \ Mpc^{-1}}$, while
studies of the Sunyaev-Zeldovich effect in galaxy clusters give a
significantly lower value $H_0=60 \pm 10\ {\rm km \ s^{-1} \
  Mpc^{-1}}$ \citep{krauss01a}.  Estimates of the ages of the oldest
globular clusters suggest $t_0 = 12.5 \pm 2.5 $ Gyrs, at the $95\%$
confidence level
\citep{krauss01,krauss01b,gnedin01,hansen02,gratton03,marchi03} and
this age estimate is consistent with several other measurements
including observations of eclipsing spectroscopic binaries
\citep{thompson01,krauss02}, results from radioactive dating of a
metal-poor star \citep{cayrel01} and WMAP data \citep{spergel03} (see
also \cite{alcaniz}). The results from the WMAP experiment suggest
$t_0 = 13.4 \pm 0.3$ Gyrs with a Hubble parameter $H_0=72 \pm 5\ {\rm
  km \ s^{-1} \ Mpc^{-1}}$, for $\l$CDM cosmology (which satisfies the
WEC). Adding SDSS and SNe Ia data to WMAP, \cite{tegmark} find an age of
$t_0 = 14.1^{1.0}_{0.9}$ Gyrs for a slightly closed $\l$CDM universe
with $H_0 = 66^{6.7}_{6.4}\ {\rm km \ s^{-1} \ Mpc^{-1}}$.  Although
these results cannot be carried over to evolving dark energy models
including those implied by our best-fit reconstruction (which violate
the WEC) they provide an indication of the range within which the age
of the universe might vary.  Keeping in mind these various results, we
use two different priors on the Hubble parameter: $H_0=72 \pm 8\ {\rm
  km \ s^{-1} \ Mpc^{-1}}$ ($1\sigma$ bound from HST; \cite{HST}), and
$H_0=66 \pm 6\ {\rm km \ s^{-1} \ Mpc^{-1}}$ (approximate bound from
WMAP, SDSS, SNe Ia; \cite{tegmark}).  For each case, we choose three
different Gaussian priors on the present age of the universe: $t_0=13
\pm 1 \ {\rm Gyrs}, \ 14 \pm 1 \ {\rm Gyrs}, \ {\rm and} \ 15 \pm 1 \
{\rm Gyrs}$ respectively, and perform the reconstruction for a
$\omt=0.3$ universe. The results are shown in the
figure~\ref{fig:age_pr}. We find that, for a Hubble parameter of
$H_0=72 \pm 8\ {\rm km \ s^{-1} \ Mpc^{-1}}$, and with an additional
prior on the age of the universe $t_0=13 \pm 1$ Gyrs, the best-fit
remains nearly the same, showing a rapid evolution of the equation of
state from $w\sim 0$ at $z \sim 1$ to $w \sim -1.2$ at $z=0$, and the
errors become narrower. As the age is increased, the best-fit equation
of state evolves more slowly, and the $\chi^2_{\rm dof}$ also
increases (see Table~\ref{tab:chi_pr}). For the prior $H_0=66 \pm 6\
{\rm km \ s^{-1} \ Mpc^{-1}}$, we find that the lowest $\chi^2_{\rm
  dof}$ is obtained for the age prior of $t_0=14 \pm 1$ Gyrs, which
once again matches our best-fit. It should be noted that the errors
are smaller in all cases, even though the $\chi^2$ may be larger. We
must remember that the addition of a new prior which is consistent
with the underlying dataset would lead to a natural reduction in
errors.  However, the addition of a prior inconsistent with the
dataset would lead to a shift of the likelihood maximum as well as a
reduction in errors, and the results would then fail to reflect the
actual information present in the dataset. That this is happening here
for the higher values of age can be seen from the fact that although
the errors are reduced, the $\chi^2_{\rm dof}$ is actually larger.
Therefore priors from other observations should be added prudently to
ensure that they do not lead to incorrect representation of the data.
Since there is as yet no clear model independent consensus on the age
of the universe, the results we obtain in this section should be
interpreted with a degree of caution.

Figure~\ref{fig:dec} shows the evolution of the deceleration parameter
with redshift.  We find that the behaviour of the deceleration
parameter for the best-fit universe is quite different from that in
$\l$CDM cosmology.  Thus, the current value of $q_0 \simeq -0.9$ is
significantly lower than $q_0 \simeq -0.55$ for $\l$CDM (assuming $\omt
= 0.3$).  Furthermore the rise of $q(z)$ with redshift is much steeper
in the case of the best-fit model, with the result that the universe
begins to accelerate at a comparatively lower redshift $z \simeq 0.3$
(compared with $z \simeq 0.7$ for $\l$CDM) and the matter dominated
regime ($q \simeq 1/2$) is reached by $z \sim 1$.

\begin{figure*}
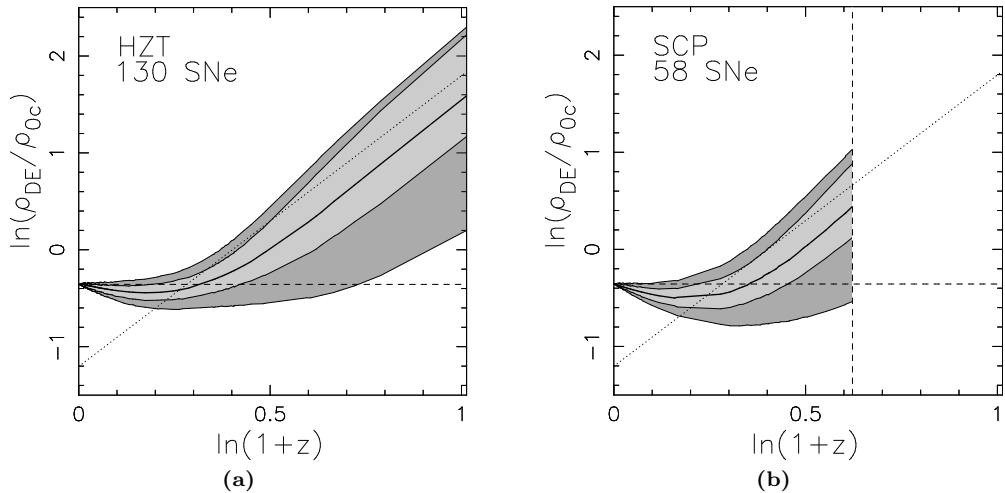

\centering
\begin{center}
\vspace{-0.05in}
$\begin{array}{c@{\hspace{0.4in}}c}
\multicolumn{1}{l}{\mbox{}} &
\multicolumn{1}{l}{\mbox{}} \\ [0.0cm]
\epsfxsize=2.4in
\epsffile{h_exp_dens_hzt.epsi} &
\epsfxsize=2.4in
\epsffile{h_exp_dens_scp.epsi} \\
\mbox{\bf (a)} & \mbox{\bf (b)}
\end{array}$
\end{center}
\caption{\small
The logarithmic variation of dark energy density $\rho_{\rm
  DE}(z)/\rhoc$ (where $\rhoc=3 H_0^2/8 \pi G$ is the present day
critical density) with redshift for $\omt=0.3$, using (a) HZT data,
and (b) SCP data.  The reconstruction is done using the polynomial fit
to dark energy, equation~(\ref{eq:taylor}). No priors are assumed on
$w(z)$.  In both panels, the thick solid line shows the best-fit, the
light grey contour represents the $1\sigma$ confidence level, and the
dark grey contour represents the $2\sigma$ confidence level around the
best-fit.  The horizontal dashed line represents $\l$CDM, and the
dotted line represents matter density.  }
\label{fig:data_dens}
\end{figure*}

\begin{figure*}
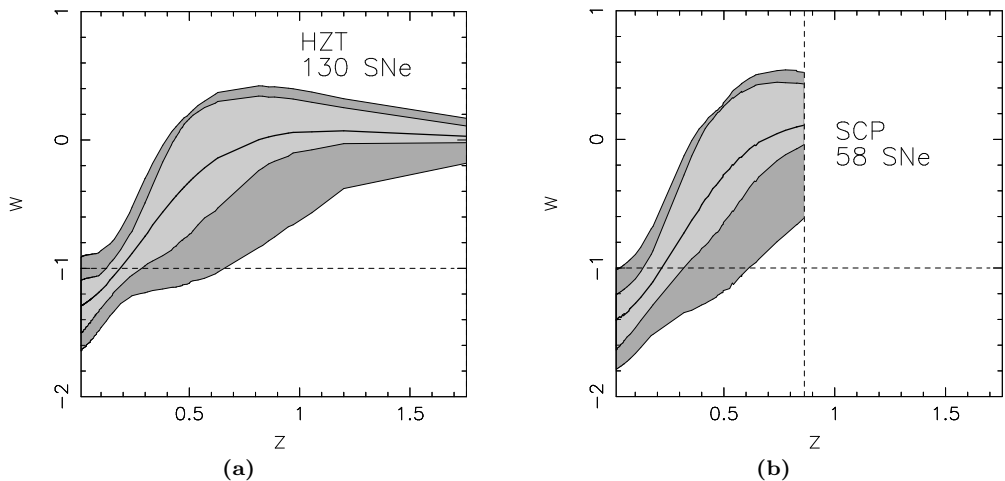

\centering
\begin{center}
\vspace{-0.05in}
$\begin{array}{c@{\hspace{0.4in}}c}
\multicolumn{1}{l}{\mbox{}} &
\multicolumn{1}{l}{\mbox{}} \\ [0.0cm]
\epsfxsize=2.4in
\epsffile{h_exp_w_hzt.epsi} &
\epsfxsize=2.4in
\epsffile{h_exp_w_scp.epsi} \\
\mbox{\bf (a)} & \mbox{\bf (b)}
\end{array}$
\end{center}
\caption{\small
The evolution of $w(z)$ with redshift for $\omt=0.3$, using (a) HZT
data, and (b) SCP data. The reconstruction is done using the
polynomial fit to dark energy, equation~(\ref{eq:taylor}). No priors are
assumed on $w(z)$.  In both panels, the thick solid line shows the
best-fit, the light grey contour represents the $1\sigma$ confidence
level, and the dark grey contour represents the $2\sigma$ confidence
level around the best-fit. The horizontal dashed line represents
$\l$CDM.  }
\label{fig:data_w}
\end{figure*}

\begin{figure*}
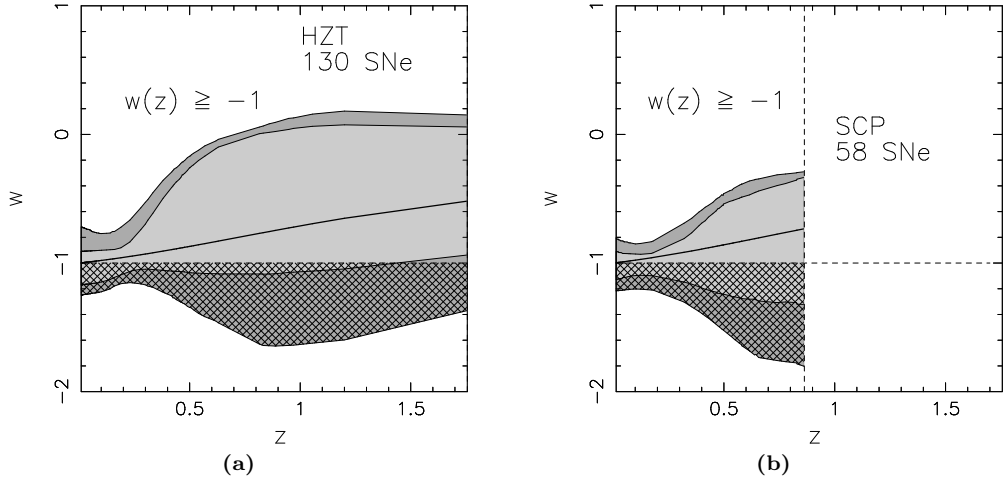

\centering
\begin{center}
\vspace{-0.05in}
$\begin{array}{c@{\hspace{0.4in}}c}
\multicolumn{1}{l}{\mbox{}} &
\multicolumn{1}{l}{\mbox{}} \\ [0.0cm]
\epsfxsize=2.4in
\epsffile{h_exp_w_hzt_cons.epsi} &
\epsfxsize=2.4in
\epsffile{h_exp_w_scp_cons.epsi} \\
\mbox{\bf (a)} & \mbox{\bf (b)}
\end{array}$
\end{center}
\caption{\small
The evolution of $w(z)$ with redshift for $\omt=0.3$, with the prior
$w(z) \geq -1, ~z \geq 0$, using (a) HZT data, and (b) SCP data.  The
reconstruction is done using the polynomial fit to dark energy,
equation~(\ref{eq:taylor}). In both panels, the thick solid line shows
the best-fit, the light grey contour represents the $1\sigma$
confidence level, and the dark grey contour represents the $2\sigma$
confidence level around the best-fit. The hatched region is forbidden
by the prior $w(z) \geq -1$. The horizontal dashed line represents
$\l$CDM.  }
\label{fig:data_c}
\end{figure*}

\subsection{Robustness of Results :}

Based on the above analysis, it is tempting to conclude that the
dominant component of the universe today is dark energy with a steeply
evolving equation of state which marginally violates the weak energy
condition.  (Of course, the less radical possibility of weakly time
dependent dark energy satisfying the weak energy condition remains an
alternative, too.) However, before any such dramatic claims are made,
we need to check if our results are in any fashion a consequence of
inherent bias in the statistical analysis itself, or in the sampling
of the data. We therefore perform the following simple exercises to
satisfy ourselves of the robustness of our results.

{\bf Using Different Subsets of Supernova Data :}

In an attempt to understand how the nature of the reconstructed
equation of state is dependent on the distribution of data, we perform
the reconstruction exercise on different samples of data. We have
confined ourselves to the case where $\omt=0.3$ for these exercises.
Firstly, we may exclude the SCP data points from the 172 SNe primary
fit, leading to a subsample of 130 SNe. We call this the HZT sample.
Figures~\ref{fig:data_dens}(a) and~\ref{fig:data_w}(a) show the result
of performing the analysis on this subsample without any constraints.
The $\chi^2$ per degree of freedom for the best-fit is $\chi^2_{\rm
  HZT}=0.9707$, which is lower than $\chi^2_{\l {\rm CDM}}=0.9939$ for
this sample. In this case we find that, though the error bars are
slightly larger, overall the dark energy density behaves in the same
way as before (compare figure~\ref{fig:data_dens}(a) with
figure~\ref{fig:dens}), showing phantom like ($w < -1$) behaviour at
lower redshifts and tracking matter at higher redshifts.  The equation
of state of dark energy also evolves much in the same way as when the
entire sample is used (compare figure~\ref{fig:data_w}(a) with
figure~\ref{fig:total}(b)), starting at $w_0 \lleq -1$ and evolving
rapidly to $w \simeq 0$.  We may also use a sample complementary to
this sample, where all the SCP data points published till date are
considered, along with the low redshift Calan-Tololo sample. This
leads to a sample of 58 SNe \citep{perl,knop}, which we call the SCP
sample.  Using this sample, we obtain the
figures~\ref{fig:data_dens}(b) and~\ref{fig:data_w}(b).  The best-fit
has a chi-squared per degree of freedom: $\chi^2_{\rm SCP}=1.0147$,
lower than $\chi^2_{\l {\rm CDM}}=1.0369$ for this sample. We find
that here too, the dark energy density initially decreases and then
starts tracking matter. The equation of state shows signs of rising
steeply at low redshifts, but since the highest redshift in this
sample is $z=0.86$, the behaviour of $w$ beyond this redshift cannot
be predicted, therefore the apparent flattening out of the curve
beyond a redshift of one cannot be seen in this case. For both these
subsets of data, we may repeat the exercise using the prior $w(z) \geq
-1 \ (z \geq 0)$. The results obtained for the equation of state, as
seen in figures~\ref{fig:data_c}(a), (b), are once again commensurate
with the results obtained earlier for the full sample
(figure~\ref{fig:constr}(b)). We may therefore conclude from this
exercise that subsampling the data does not significantly affect our
results, and the steep evolution of the equation of state of dark
energy is not a construct of the uneven sampling of the supernovae,
but rather, reflects the actual nature of dark energy.

{\bf Testing our Ansatz against fiducial dark energy models :}

\begin{figure*}
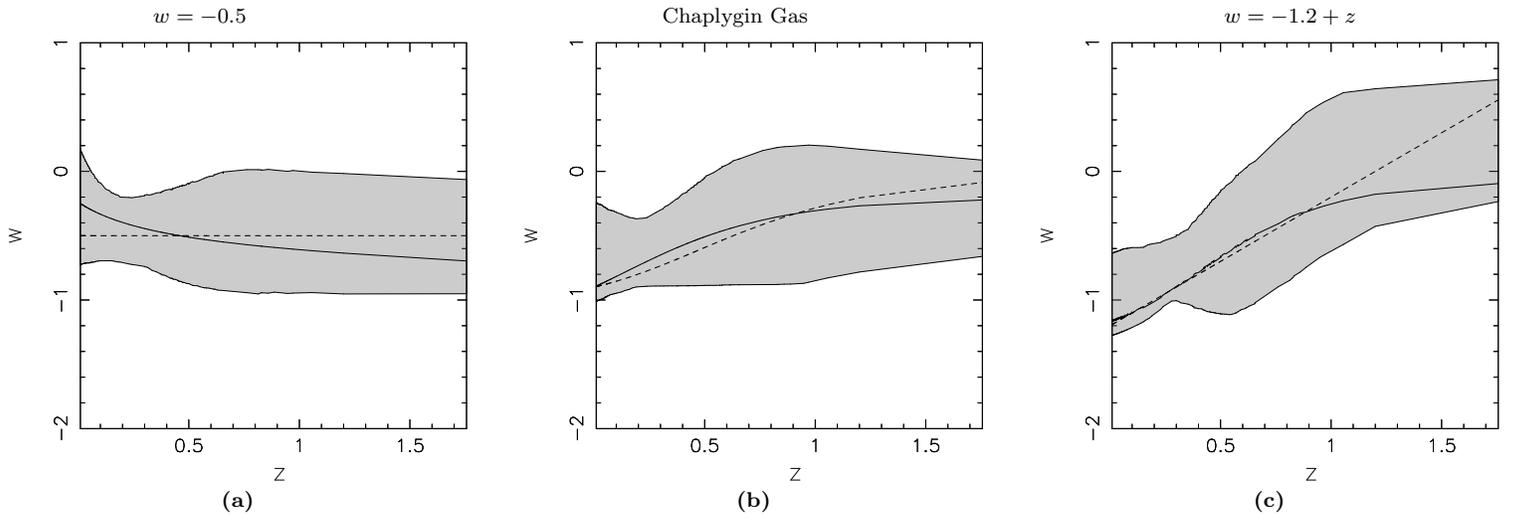

\centering
\begin{center}
\vspace{-0.05in}
\centerline{\mbox{\hspace{0.in} $w=-0.5$ \hspace{2.1in} ${\rm Chaplygin \ Gas}$ \hspace{2.1in} $w=-1.2+z$}}
$\begin{array}{@{\hspace{-0.4in}}c@{\hspace{0.3in}}c@{\hspace{0.3in}}c}
\multicolumn{1}{l}{\mbox{}} &
\multicolumn{1}{l}{\mbox{}} &
\multicolumn{1}{l}{\mbox{}} \\ [-0.5cm]
\epsfxsize=2.4in
\epsffile{h_exp_w_sim_w.epsi} &
\epsfxsize=2.4in
\epsffile{h_exp_w_sim_cg.epsi} &
\epsfxsize=2.4in
\epsffile{h_exp_w_sim_ev.epsi} \\
\mbox{\bf (a)} & \mbox{\bf (b)} & \mbox{\bf (c)}
\end{array}$
\end{center}
\caption{\small
Reconstructed equation of state, $w(z)$, for simulated data
corresponding to three fiducial dark energy models: (a) quiessence
with $w=-0.5$, (b) generalised Chaplygin gas: $p=A/\rho^{\alpha}$,
with $\alpha=0.5, w_0=-0.9$, and (c) $w=w_0+w_1 z$ with $w_0=-1.2,
w_1=1.0$.  $\omt=0.3$ is assumed and the reconstruction is done using
the polynomial fit to dark energy, equation~(\ref{eq:taylor}). In each
panel, the thick solid line is the best-fit, the dashed line
represents the exact model value, and the light grey contour
represents the $1\sigma$ confidence level around the best fit.  }
\label{fig:sim_err}
\end{figure*}

\begin{figure*}
\centering
\begin{center}
\vspace{-0.05in}
$\begin{array}{c@{\hspace{0.4in}}c}
\multicolumn{1}{l}{\mbox{}} &
\multicolumn{1}{l}{\mbox{}} \\ [0.0cm]
\epsfxsize=2.4in
\epsffile{lin_w.epsi} &
\epsfxsize=2.4in
\epsffile{lin_w_cons.epsi} \\
\mbox{\bf (a)} & \mbox{\bf (b)}
\end{array}$
\end{center}
\caption{\small
Evolution of $w(z)$ with redshift for $\omt=0.3$ with (a) no priors on
$w(z)$, and (b) the prior $w(z) \geq -1 \ (z\geq 0)$. The
reconstruction is done using Linder's fit to the equation of state,
equation~\ref{eq:lin}. In both panels, the thick solid line represents
the best-fit, the dashed line represents $\l$CDM, and the light
grey contour represents the $1\sigma$ confidence level. In the right
hand panel, the hatched region is forbidden by the prior $w(z) \geq
-1$.}
\label{fig:lin}
\end{figure*}

\begin{figure*}
\centering
\begin{center}
\vspace{-0.05in}
$\begin{array}{c@{\hspace{0.4in}}c}
\multicolumn{1}{l}{\mbox{}} &
\multicolumn{1}{l}{\mbox{}} \\ [0.0cm]
\epsfxsize=2.4in
\epsffile{bass_w.epsi} &
\epsfxsize=2.4in
\epsffile{bass_w_cons.epsi} \\
\mbox{\bf (a)} & \mbox{\bf (b)}
\end{array}$
\end{center}
\caption{\small
Evolution of $w(z)$ with redshift for $\omt=0.3$ with (a) no priors on
$w(z)$, and (b) the prior $w(z) \geq -1 \ (z\geq 0)$. The
reconstruction is done using a four-parameter fit to the equation of state,
equation~\ref{eq:bass}. In both panels, the thick solid line represents
the best-fit, the dashed line represents $\l$CDM, and the light
grey contour represents the $1\sigma$ confidence level. In the right
hand panel, the hatched region is forbidden by the prior $w(z) \geq
-1$.}
\label{fig:bass}
\end{figure*}

The crucial question of course is whether the reconstructed equation
of state of dark energy depends upon the ansatz which is used in the
exercise, \ie, whether the behaviour of the equation of state merely
reflects a bias in the ansatz itself. In this section we show how the
ansatz performs in recovering dark energy models whose equation of
state is known, from simulated data.  This ansatz was demonstrated to
work extremely well when simulations of SNAP data were used
\citep{alam03a}. However, simulation of SNAP-like data is an
optimistic exercise, since data of this quality is unlikely to be
available in the near future. We now demonstrate the accuracy with
which the ansatz can recover the fiducial background cosmological
model if data is simulated using present-day observational standards.
In figures~\ref{fig:sim_err} (a), (b), (c), we show how well the
ansatz recovers the equation of state for three fiducial models
(assuming $\omt=0.3$):

(a) a quiessence dark energy model with a constant equation of state:
$w=-0.5$,

(b) a generalised Chaplygin gas model with $p=A/\rho^{\alpha}$: with
$\alpha=0.5$ and the present-day equation of state $w_0=-0.9$, which
would give rise to an effective equation of state
\beq
w(z)=- \frac{|w_0|}{|w_0|+(1-|w_0|) (1+z)^{3 (1+\alpha)}} \,,
\eeq
and

(c) a model with a linearly evolving equation of state: $w(z)=w_0+w_1
z$, with $w_0=-1.2, w_1=1$.

(For DE models with $w = -1,-2/3,-1/3$ the ansatz is exact therefore
we don't show the results for these cases.)

We find that in all three cases, the fiducial model lies within the
$68 \%$ confidence limits around the best-fit $w(z)$.  Based on this
result, we claim that within the $1\sigma$ error bars, the
reconstructed equation of state represents the true properties of dark
energy when we use real data.

{\bf Using other Ansatz :}

It is also important to check whether the results of our
reconstruction can be replicated using other ansatz such as fits to
the luminosity distance or the equation of state. Many different fits
have been suggested in the literature (see for example \cite{turner},
\cite{saini00}, \cite{albrecht}, \cite{efstathiou}).  Here we choose the
fit suggested in \cite{linder} in which the equation of state of dark
energy is expanded as
\beq
w(z)=w_0+\frac{w_1 z}{1+z}~.
\label{eq:lin}
\eeq
The luminosity distance can therefore be expressed as
\beq
\frac{d_L(z)}{1+z} = \frac{c}{H_0} \int_1^{1+z} \frac{dx}{\sqrt{\omt x^3+\omx}} \,,\\
\eeq
where $\omx = (1-\omt) x^{3 (1+w_0+w_1)} {\rm exp}[3 w_1
(\frac{1}{x}-1)]$.

We find that for this fit, the errors in the equation of state get
larger with redshift, however this fit too demonstrates that the
equation of state of dark energy increases rapidly with redshift
(figure~\ref{fig:lin}(a)) when no priors are assumed on the equation
of state (EOS). The $\chi^2$ per degree of freedom at the best-fit is
$\chi^2_{\rm dof}=1.0298$. When the prior $w(z) \geq -1 \ (z \geq 0)$
is invoked, the best-fit EOS remains very close to the $\l$CDM
model (figure~\ref{fig:lin}(b)).  Therefore, from this ansatz, we may
make the statement that at low redshifts, the equation of state of
dark energy shows the same signs of rising steeply with redshift if no
priors are assumed on the equation of state, thus supporting our
earlier results. The large errors in the equation of state at
redshifts of $z \ggeq 0.5$ however make it difficult to make any
definitive statements about the behaviour of dark energy at high
redshifts.

A limitation of the fit (\ref{eq:lin}) is that it is unable to describe
very rapid variations in the equation of state. An ansatz
which accommodates this possibility has been 
suggested in \cite{bassett}
\beq
w(z)=w_i+\frac{w_f-w_i}{1+{\rm exp}(\frac{z-z_t}{\Delta})}\,\,,
\label{eq:bass}
\eeq
where $w_i$ is the initial equation of state at high redshifts, 
$z_t$ is a transition redshift at which the equation of state falls to
$w(z_t) = (w_i + w_f)/2$ and $\Delta$ describes the rate of change of $w(z)$.

The resulting luminosity distance has the form:
\beq
\frac{d_L(z)}{1+z} = \frac{c}{H_0} \int_1^{1+z} \frac{dx}{\sqrt{\omt x^3+\omx}} \,,\\
\eeq
where $\omx = (1-\omt) {\rm exp}[3 \int_0^{x-1} (1+w(z)) dz/(1+z)]$.

The results for the analysis using this fit to the equation of state
are shown in fig.~\ref{fig:bass}. We find that when the reconstruction
is done without any priors on the equation of state
(figure~\ref{fig:bass}(a)), the best fit is {\em remarkably close} to the
result for ansatz~(\ref{eq:taylor}) (figure~\ref{fig:total}(b)). The
$\chi^2$ per degree of freedom at the minimum is $\chi^2_{\rm
  dof}=1.0175$ for this fit. The errors in this case are somewhat
larger, especially at high redshift. If we constrain $w(z) \geq -1$,
then as before, the evolution of the equation of state is much slower
(figure~\ref{fig:bass}(b)). So the reconstruction using this ansatz
appears to confirm our earlier results.

The above exercises lead us to conclude that our results are neither
dependent on the nature of the statistical analysis nor on the manner
in which the SNe data is sampled. It therefore appears that dark energy
with a steeply evolving equation of state provides a compelling
alternative to a cosmological constant if data are analysed in a
prior-free manner and the weak energy condition is not imposed by
hand.

\subsection{Reconstructing dark energy using a new Supernova sample}

\begin{figure}
\centering
\begin{center}
\vspace{-0.05in}
\psfig{figure=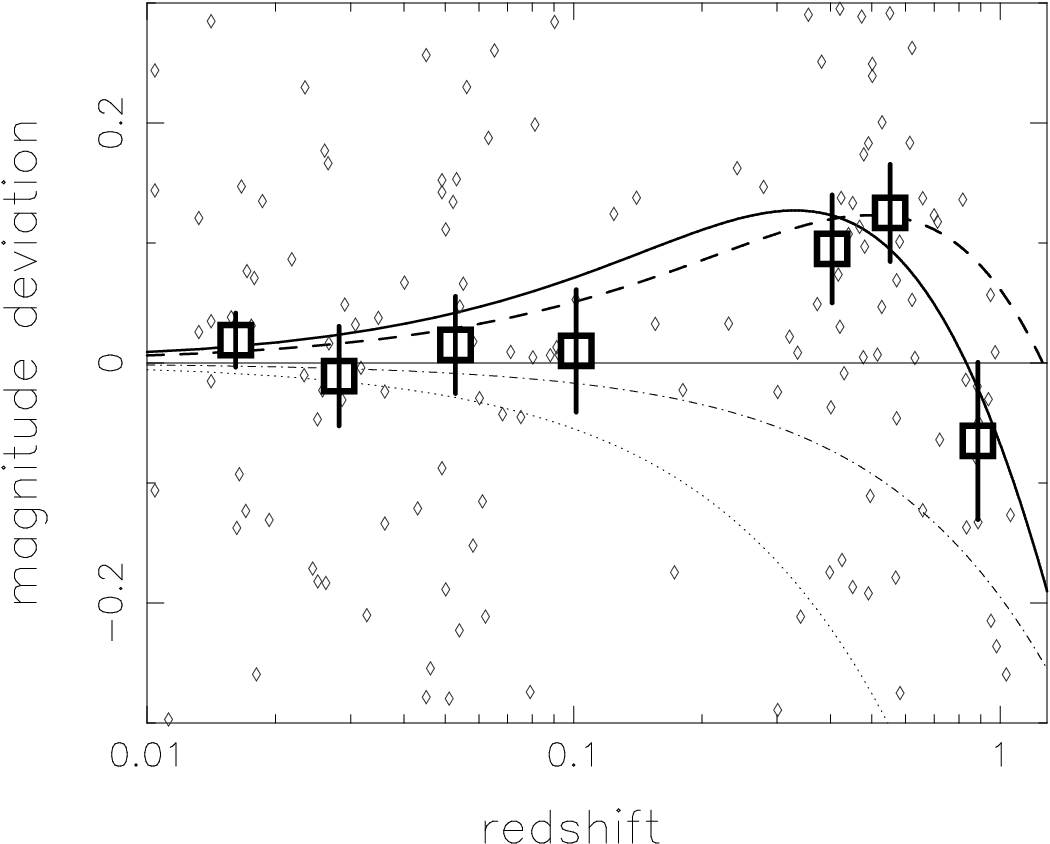,width=0.4\textwidth,angle=0}
\end{center}
\caption{\small
Literature supernovae (diamonds) shown along with median values binned
by redshift (large squares). Individual points are shown without error
bars for clarity. The solid horizontal line represents the empty
universe with $(\omt, \ol)=(0.0,0.0)$. The thick solid line represents
the magnitude deviation of our best-fit universe for this data-set
from the empty universe. The thick dashed line represents $\l$CDM with
$(\omt, \ol)=(0.3,0.7)$, the dot-dashed and dotted lines represent
cosmologies with $(\omt, \ol)=(0.3,0.0) \ {\rm and} \ (1.0,0.0)$
respectively.}
\label{fig:hubble}
\end{figure}

\begin{figure}
\centering
\begin{center}
\vspace{-0.05in}
\epsfxsize=2.4in
\epsffile{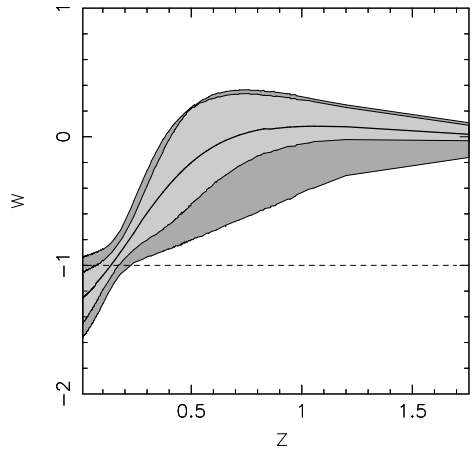}
\end{center}
\caption{\small
The evolution of $w(z)$ with redshift for $\omt=0.3$. The
reconstruction is done using the polynomial fit to dark energy,
equation~(\ref{eq:taylor}), for the latest HZT sample of 194 type Ia
SNe. The thick solid line shows the best-fit, the light grey contour
represents the $1\sigma$ confidence level, and the dark grey contour
represents the $2\sigma$ confidence level around the best-fit. The
dashed line represents $\l$CDM. No priors are assumed on $w(z)$.
}
\label{fig:w_new}
\end{figure}

As this paper was nearing completion, a new dataset consisting of 23
type Ia SNe was released by the HZT team \citep{hzt_new}. It is
clearly important to check whether or not these new data points
corroborate the findings reported in the previous sections.
Accordingly, we use a subset of 200 type Ia SNe with $A_V \leq 0.5$
from the 230 SNe sample of \cite{tonry03}, and 22 SNe with $A_V \leq
0.5$ from the new sample to obtain a best-fit for our ansatz with
$\omt=0.3$. We then plot the magnitude deviation of our best-fit
universe from an empty universe with $(\omt,\ol)=(0.0,0.0)$ in order to
illustrate how well our model fits the data (figure~\ref{fig:hubble}).
For clarity, we plot the median values of the data points. We obtain
medians in redshift bins by requiring that each bin has a width of at
least 0.25 in log$z$ and contain at least 20 SNe. For comparison, we
also plot an $\l$CDM $(\omt,\ol)=(0.3,0.7)$ model, as well as OCDM and
SCDM models.  From figure~\ref{fig:hubble} we see that our dark energy
reconstruction is a much better fit to SNe beyond $z \sim 0.8$ than
$\l$CDM.  At low redshifts ($z \sim 0.1$) the agreement between data and
the two models is rather marginal.  We now add 22 of the new
supernovae (rejecting one with $A_V > 0.5$) to our existing dataset of
172 supernovae and perform DE reconstruction on this new dataset of
194 SNe, assuming $\omt=0.3$ and no other priors. The resultant
figure~\ref{fig:w_new} is similar to the figure~\ref{fig:total}(b),
with slightly smaller errors and has a best-fit $\chi^2_{\rm
  dof}=1.015$.  The above exercises point to the robustness of results
reported in previous sections, and indicate that evolving dark energy
agrees well with the full data set containing 194 type Ia SNe.

\section{\bf The accelerating universe and the {\em Energy Conditions}}

The energy conditions :

\begin{itemize}

\item Strong energy condition: $\rho + 3p \geq 0$ (SEC),

\item Weak energy condition: $\rho \geq 0$, $\rho + p \geq 0$ (WEC)

\end{itemize}
play a vitally important role in our understanding of the accelerating
universe, both in the context of inflation and dark energy. We
therefore consider it worthwhile to review certain key developments
which deepened our understanding of these issues.

In an expanding FRW universe the SEC implies that the universe
decelerates while the WEC forbids the pressure from becoming too
negative. Additionally, in the 1960's and early 1970's it was noted
that energy conditions play a crucial role in the formulation of the
singularity theorems in cosmology.  Indeed, one of the necessary
conditions for the existence of an initial/final singularity in big
bang cosmology is that matter satisfies both the SEC and WEC
\citep{hawellis}.

By the late 1970's it became clear that not all forms of matter
satisfy the energy conditions. Perhaps the best example of a form of
matter which satisfied the weak energy condition but violated the
strong one is the cosmological constant, introduced into cosmology by
Einstein in 1917.  In addition, the vacuum expectation value of the
energy momentum tensor, $\langle T_{ik}\rangle_{\rm vac}$, which
describes quantum effects (particle production and vacuum
polarization) in an expanding universe could, in certain cases,
violate both WEC and SEC \citep{bd82,gmm80}.  (For certain
space-times, such as de Sitter space, the vacuum energy momentum
tensor generates a cosmological constant since $\langle
T_{ik}\rangle_{\rm vac} = \l g_{ik}$.)  Thus by the late 1970's it was
well known that neither of the energy conditions could be held as
being sacrosanct.\footnote{The importance of quantum effects to dark
  energy model building has been emphasised in \cite{sh98,parker99}.}

The 1980's, as we all know, led to great advances in the development
of the inflationary paradigm. The inflaton field mimics the behaviour
of a cosmological constant over sufficiently small intervals of time
and therefore violates the SEC.  Early dark energy models were based
on inflaton-type scalars which coupled minimally to gravity
(quintessence). Quintessence violates the SEC but respects the WEC.
Precisely because of the latter property, not {\em any} experimentally
obtained $d_L(z)$ is compatible with quintessence, as emphasized in
\cite{ss00}.  (The same observation holds for $H(z)$, since the latter
can be derived from $d_L(z)$ using equation (\ref{eq:H}).)  Clearly if
observations do indicate that the WEC is violated by DE then more
general (WEC-violating) models for DE should be seriously considered.
One example of WEC-violating DE is provided by scalar-tensor gravity.
Scalar-tensor models contain at least two functions of the scalar
field (dilaton) describing dark energy.  As shown in \cite{beps00},
these two functions, namely, the scalar field potential and its
coupling to the Ricci scalar $R$, are sufficiently general to explain
{\em any} $H(z)$ obtained from observations.

The WEC can also be effectively violated in DE models constructed in
braneworld cosmology.  It has recently been shown that such models,
with $w_{\rm eff} < -1$, are in excellent agreement with supernova
data \citep{as02}.  Since the field equations in these models are
derived from a higher dimensional Lagrangian the unusually rapid
acceleration of the four dimensional universe arises because of the
full five dimensional theory and not because of matter which continues
to satisfy the energy conditions and whose density remained finite and
well behaved at all times \citep{ss02a}.  This behaviour is in
contrast to phantom, which assumes a conventional `perfect fluid' form
for the energy-momentum tensor and therefore contains pathological
features such as an energy density which diverges in the future and a
sound speed which is faster than that of light \citep{caldwell,ss02a}.

The fact that the observed luminosity distance (derived from supernova
observations) is better fit by dark energy violating the WEC than
either quintessence or a cosmological constant was first noticed by
\cite{caldwell}.  Caldwell called this `phantom energy' and showed
that larger values of $\omt$ ($\omt \ggeq 0.2$) implied increasingly
more negative values for the equation of state ($w \lleq -1$) of
phantom. \footnote{Note, however, that we will not consider the
  theoretical model of phantom matter based on a ghost scalar field
  proposed in this paper since, as is well known, it is unstable with
  respect to particle creation (particle + antiparticle of the ghost
  scalar field plus particle + antiparticle of all usual matter
  fields) and to the loss of spatial homogeneity at both quantum and
  non-linear classical levels.}
Caldwell's results have since been confirmed by larger and better
quality SNe data sets -- for instance \cite{knop} find that, in the
absence of priors being placed on $\omt$, the DE equation of state has
a $99\%$ confidence limit of being $< -1$ !  Both \cite{caldwell} and
\cite{knop} however work under the assumption that the equation of
state of dark energy is unevolving, so that $w = $ constant.

In this paper we have shown that, suspending the WEC prior and
allowing the dark energy equation of state to evolve brings out
dramatically new properties of dark energy.  Thus the dark energy
model which best fits the SNe observations has an equation of state
which rapidly evolves from $w(z) \lleq -1$ at present ($z = 0$) to
$w(z) \simeq 0$ at $z \sim 1$.  Dark energy therefore appears to have
properties which interpolate between those of dark matter (dust) at
early times and those of a `phantom' ($w \lleq -1$) at late times.

\section{\bf Conclusions and discussion}
This paper reports the model independent reconstruction of the cosmic
equation of state of dark energy in which no priors are imposed on
$w(z)$.  In the literature the imposition of various priors frequently
precedes the analysis of observational data sets. Such a procedure is
well founded and entirely justified when priors are dictated by
complementary information such as orthogonal observations coming from
different data sets.  However, on occasion the use of priors is
justified on `theoretical grounds' and in this case one must be
careful so as not to prejudge nature.  (Compelling theoretical reasons
might well reflect our own particular conditioning or set of
prejudices !)  In the case of the analysis of type Ia supernova data,
the priors most frequently used have been $w = $ constant and $w \geq
-1$.  Both confine DE to within a narrow class of models. Moreover, as
shown in \cite{maor02}, the imposition of such priors on the cosmic
equation of state can, on occasion, lead to gross misrepresentations
of reality.

In this paper we do not impose any priors on $w(z)$ and reconstruct
the equation of state of dark energy in a model independent manner. In
this case our best fit $w(z)$ evolves from $w \lleq -1$ at $z=0$, to
$w \simeq 0$ at $0.8 \lleq z \lleq 1.75$ (the upper limit is set by
observations).  This result is robust to changes in the value of
$\omt$ and remains in place within the broad interval $ 0.1 \leq \omt
\leq 0.5$.  Our reconstruction clearly favours a model of DE whose
equation of state metamorphoses from $w=0$ in the past to $w \simeq
-1$ today.  An excellent example of a model which has this property is
the Chaplygin gas \citep{chap1}.  However, in this model dark energy
does not violate the weak energy condition (if it was not already
violated initially). Our results also lend support to the dark energy
models discussed in \cite{bassett,corasaniti} in which the DE equation
of state shows a late-time phase transition. An interesting example of
an evolving DE model in which $w_0 < -1$ at present whereas $w(z) >
-1$ at earlier times is provided by the braneworld models (called
BRANE1) examined in \cite{ss02a} which have been shown to agree very
well with current supernova observations \citep{as02}.

It is also conceivable that the observed rapid growth in the EOS might
characterise `unified' models of dark matter (DM) and dark energy
(DE).  We end this paper with a small speculation on this last
possibility.  Since the nature of both DM and DE is currently unknown,
it may be that a mechanism exists which converts DM (with $w=0$) into
DE (with $w \simeq -1$) in regions with sufficiently high density
contrast $\delta\rho/\rho \gg 1$.  (This would happen if, for
instance, the rate of conversion of DM into DE depended upon
$(\delta\rho/\rho)^x, x \gg 1$, etc.)  Since the conversion of DM to
DE is confined to high peaks of the density field this process will
not occur uniformly in the entire universe but will be restricted to
regions occupying a small filling fraction ($FF$) ($FF \ll 1$ for
regions with $\delta\rho/\rho \gg 1$; see for instance \cite{sheth03}
and references therein).  This process could commence as early as $z
\sim 10 - 20$ when the first peaks in a CDM model collapse.  Since DE
does not cluster and since $\rho_{\rm DE}/\rho_{\rm DM}$ grows rapidly
as the universe expands, DE from high density regions ($FF \ll 1$)
will spread at the speed of light, percolating through the entire
universe ($FF \sim 1$) by $z \sim 1$.  Since the creation of DE is
tagged to the formation of structure, this model may not encounter the
`coincidence problem' which plagues other scenarios of DE including
quintessence. (However this model might have problems in producing a
sufficiently homogeneous and isotropic distribution of dark energy on
the largest scales.)  The concrete mathematical framework for a
phenomenological model of this kind will be worked out in a companion
paper.

In summary, evolving DE models have been shown to satisfy SNe
observations just as well (if not better) than the cosmological
constant.  Our best fit equation of state, in the absence of any
priors, evolves from $w(z) \lleq -1$ at $z = 0$ to $w(z) \simeq 0$ at
$z \sim 1$.  Indeed, figure~\ref{fig:hubble} shows that our best fit
EOS is better able to account for the relative brightness of
supernovae at $z \ggeq 0.8$ than $\l$CDM.  However, the evolution in
$w(z)$ is much weaker if the prior $w(z)\ge -1$ ($z \geq 0$) is
imposed.  Due to the paucity of SNe data beyond $z = 1.2$ (till date,
there is only a single data point beyond $z=1.2$, SN1999bf at
$z=1.75$) it is not clear whether $w(z) \simeq 0$ is a stable
asymptotic value for the reconstructed DE equation of state at high
redshifts.\footnote{An alternative explanation for the relative
  brightness of SNe at these redshifts, say, by gravitational lensing
  \citep{barber00} could clearly alter the high-z properties of our
  best-fit.}  New supernova data at $z \ggeq 1$ from ongoing as well
as planned surveys (SNAP) combined with data from other cosmology
experiments (CMB, LSS, S-Z survey's, lensing, etc.)  are bound to
provide important insights on the nature of dark energy at high
redshifts.  Our results clearly throw open exciting new possibilities
for dark energy model building.

\[ \]
\medskip
\noindent{\it Acknowledgments:}

We would like to thank John Tonry for several important clarifications
and for help in preparing figure~\ref{fig:hubble}.  We also thank
Sarah Bridle, Pier-Stefano Corasaniti, Alessandro Melchiorri, Yuri
Shtanov and Lesha Toporenskii for their useful comments on an earlier
version of the paper.  One of us (VS) acknowledges useful discussions
with Salman Habib and Daniel Holz.

UA thanks the CSIR for providing support for this work.  AS was
partially supported by the Russian Foundation for Basic Research,
grant 02-02-16817, and by the Research Program ``Astronomy'' of the
Russian Academy of Sciences.

\appendix

\section{Propagation of errors }\label{sec:errors}

We have seen that the error bars on $w(z)$ for the analysis using
ansatz~(\ref{eq:taylor}) are non-monotonic with redshift. Low redshift
behaviour of the equation of state affects the luminosity distance at
all higher redshifts, while high redshift behaviour effects fewer such
distances. This leads to an expectation that high-$z$ behaviour of the
equation of state should be poorly constrained as opposed to the
low-$z$ behaviour. This seems to contradict the behaviour seen in our
figures. To investigate if this could be explained by our specific
method of error analysis we describe the Fisher matrix error bars
below and show that they are almost identical to what we obtain in our
method.

In an analysis which uses an ansatz with $n$ parameters $p_i$, the
Fisher information matrix is defined to be 
\beq
F_{ij} \equiv \left \langle \frac{\partial^2 {\bf L}}{\partial p_i \partial p_j} \right \rangle\,\,,
\eeq
where ${\bf L}=-{\rm log}\cal{L}$, $\cal{L}$ being the likelihood .
For an unbiased estimator, the errors on the parameters will follow
the Cram\'{e}r-Rao inequality : $\Delta p_i \geq 1/\sqrt{F_{ii}}$.

Since the likelihood function is approximately Gaussian near the
maximum likelihood (ML) point, the covariance matrix for a maximum
likelihood estimator is given by
\beq
(C^{-1})_{ij} \equiv \frac{\partial^2 {\bf L}}{\partial p_i \partial p_j}\,\,.
\eeq
The Fisher information matrix is therefore simply the expectation
value of the inverse of the covariance matrix at the ML-point.
  
Given the covariance matrix, the error on any cosmological quantity
$Q(p_i)$ is given by :
\beq\label{eq:propagate}
\sigma_Q^2 = \sum_{i=1}^n \left( \frac{\partial Q}{\partial p_i} \right)^2 C_{ii}+2 \sum_{i=1}^n \sum_{j=i+1}^n  \left( \frac{\partial Q}{\partial p_i} \right)  \left( \frac{\partial Q}{\partial p_j} \right) C_{ij}\,\,.
\eeq
Thus the nature of the errors on a quantity will depend essentially on
the manner in which it is related to the parameters of the system.

We now consider how errors propagate for different cosmological
quantities for the polynomial fit to dark energy which we have used
for most of the results in this paper :
\beq
H^2/H_0^2 = \omt x^3+A_0+A_1 x+A_2 x^2,~x=1+z \,\,,
\eeq
where $A_0=1-\omt-A_1-A_2$. If $\omt$ is held constant then the
parameters of the system are $(A_1,A_2)$.

\begin{figure}
\centering
\begin{center}
\vspace{-0.05in}
\epsfxsize=2.4in
\epsffile{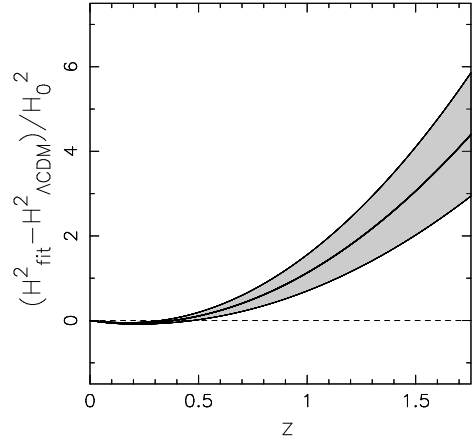}
\end{center}
\caption{\small
The deviation of $H^2/H_0^2$ from corresponding $\l$CDM values over
redshift for the ansatz~(\ref{eq:taylor}). The thick solid line shows
the best-fit and the light grey contour represents the $1\sigma$
confidence level around the best-fit.  The dashed horizontal line
denotes $\l$CDM. $\omt = 0.3$ is assumed.}
\label{fig:err_hubb}
\end{figure}

\begin{figure}
\centering
\begin{center}
\vspace{-0.05in}
\epsfxsize=2.4in
\epsffile{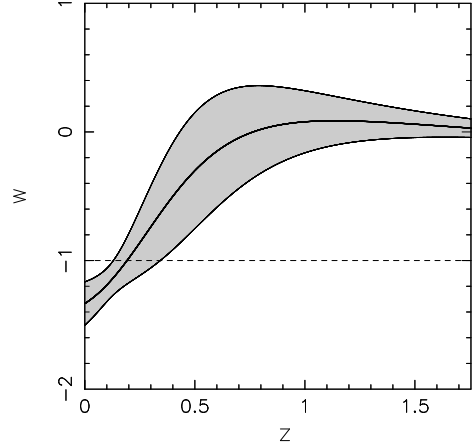}
\end{center}
\caption{\small
The variation of the equation of state of dark energy $w(z)$ over
redshift for the ansatz~(\ref{eq:taylor}). The thick solid line shows
the best-fit and the light grey contour represents the $1\sigma$
confidence level around the best-fit.  The dashed horizontal line
denotes $\l$CDM. $\omt = 0.3$ is assumed.}
\label{fig:err_w}
\end{figure}

We obtain the covariance matrix in $(A_1,A_2)$ from the ML analysis,
and then using equation~(\ref{eq:propagate}), calculate the errors
on cosmological quantities of interest. For example, the errors on the
quantity $\Delta H^2 =(H^2-H^2_{\l CDM})/H_0^2$ are given by :
\beq
\sigma_{\Delta H^2}^2(x) = (x-1)^2 [C_{11}+2 (x+1) C_{12}+(x+1)^2 C_{22}]\,\,.
\eeq

Although the term $C_{12}$ is negative we find that $\sigma_{\Delta
  H^2}^2(x)$ still {\em increases} with redshift.  This is shown in
the figure~\ref{fig:err_hubb}. The errors shown are approximately
similar to those obtained in figure~\ref{fig:dev}.

The corresponding errors on the equation of state can be calculated
using equations~(\ref{eq:state2}) and~(\ref{eq:propagate}), and has the
somewhat more complicated expression :
\beq
\sigma_w^2(x) = \frac{x^2 [f_1^2 C_{11} + 2 f_1 f_2 C_{12}+ f_2^2 C_{22}]}{9 [1-\omt+A_1 (x-1)+A_2 (x^2-1)]^4}\,\,,
\eeq
where 
\begin{eqnarray*}
f_1 &=& 1-\omt-A_2 (x-1)^2 ~, \\
f_2 &=& 2 x (1-\omt)+A_1 (x-1)^2 \,\,,
\end{eqnarray*}
and $A_1,A_2$ are the mean values of the parameters.  Although in this
case it is difficult to predict the behaviour of error bars, after
substituting the numerical values we obtain the error bars that are
shown in figure~\ref{fig:err_w}.  This figure can be compared to the
figure~\ref{fig:total}(b), having almost identical errors.

This shows that the nature of our error bars is not an artifact of our
specific method of error analysis. However, as shown in
figure~\ref{fig:lin}, a two parameter expansion in $w(z)$ shows
monotonically deteriorating errors in $w(z)$ with the redshift, while
the expansion in $H_2(z)$ shows errors that improve with redshift
(figure~\ref{fig:total}(b)). This indicates that the nature of error
bars might be affected by which quantity is being approximated. In the
limit of infinite terms in the expansion of various quantities all the
methods should produce identical result.  The practical need for
truncating these expansions make these approximations slightly
different from each other. More specifically, we require setting of
priors
\ber
f(z) &=& \sum_{i=0}^{\infty} a_n z^n\\
a_n &=& 0 ; \ (n> N_p)
\eer
where $f(z)$ could be $H(z)$, $w(z)$ or any other physical quantity
and $N_p$ is the chosen number of parameters. The non-linear priors in
the above equation make different finite expansions inequivalent.
Since we do not know for certain if the underlying model for the
accelerating expansion involves an energy component with negative
pressure in a FRW setting we are forced to choose one of the
alternatives for approximations. We hope that with increasingly high
quality data the effect of such truncations will eventually disappear.

\section{Reconstruction of Other Dark Energy Models }\label{sec:fits}

\begin{figure*}
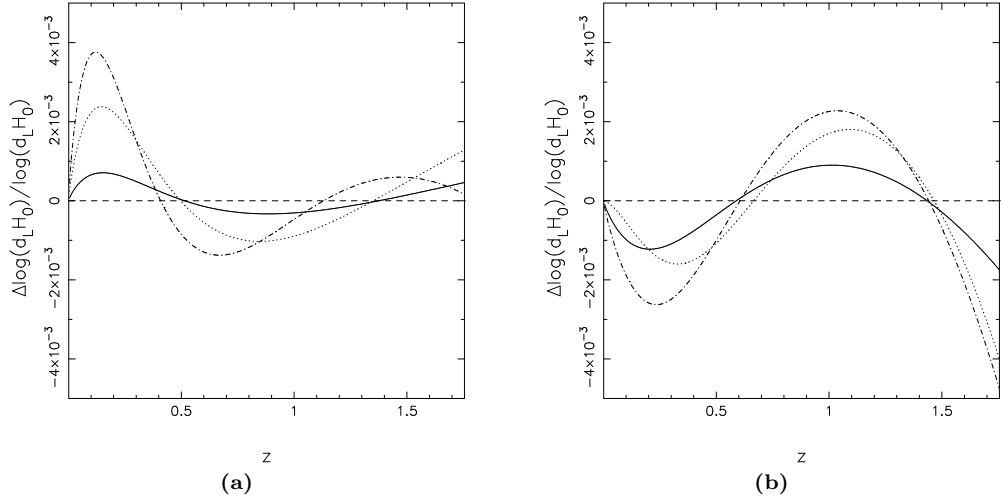

\centering
\begin{center}
\vspace{-0.05in}
$\begin{array}{c@{\hspace{0.4in}}c}
\multicolumn{1}{l}{\mbox{}} &
\multicolumn{1}{l}{\mbox{}} \\ [0.0cm]
\epsfxsize=2.4in
\epsffile{h_exp_error_lin.epsi} &
\epsfxsize=2.4in
\epsffile{h_exp_error_coras.epsi} \\
\mbox{\bf (a)} & \mbox{\bf (b)}
\end{array}$
\end{center}
\caption{\small
The deviation $\Delta{\rm log}(d_L H_0)/{\rm log}(d_L H_0)$ between
actual value and that calculated using the ansatz~(\ref{eq:taylor})
over redshift for different values of parameters for (a) the Linder
fit, and (b) the Corasaniti fit for equation of state of dark energy,
with $\omt=0.3$. In panel (a), the solid line shows the deviation for
the Linder fit with $w_0=-0.8, w_1=0.5$, the dotted line for
$w_0=-1.0, w_1=1.0$, and the dot-dashed line for $w_0=-1.2, w_1=2.0$.
In panel (b), the solid line represents the Corasaniti fit with
$w_Q^0=-0.8, w_Q^m=-0.5, z_c^m=1.5, \Delta=0.1$, the dotted line shows
$w_Q^0=-1, w_Q^m=-0.2, z_c^m=0.6, \Delta=0.07$, and the dot-dashed
line, $w_Q^0=-1.2, w_Q^m=0.1, z_c^m=0.11, \Delta=0.03$.  The dashed
horizontal line in both panels represents zero deviation from model
values, which is true for $\l$CDM, and $w=-1/3, w=-2/3$ quiessence
models.  }
\label{fig:sim}
\end{figure*}

We have seen in the figure~\ref{fig:err} that the ansatz~(\ref{eq:taylor})
works well for several physically motivated models of quintessence,
Chaplygin gas and SUGRA. In this section we take this exercise further
and see how well it can reconstruct some of the other fits to dark
energy known in literature.  In figures~\ref{fig:sim} (a) and (b), we
show results for simulations using $\omt=0.3$ and two different fits
to the equation of state of dark energy :

(a) The fit suggested in \cite{linder} : $w(z)=w_0+w_1 z/(1+z)$. For
this we consider three sets of values in order of increasing evolution
of $w(z)$ : (a) $w_0 = -0.8, w_1 = 0.5$, (b) $w_0 = -1.0, w_1 = 1.0$,
and (c) $w_0 = -1.2, w_1 = 2.0$, and,

(b) The non-perturbative $w(z)$ suggested in \cite{copeland} and
\cite{corasaniti}, which has the parameters $w_Q^0$ (the dark energy
equation of state today), $w_Q^m$ (the dark energy equation of state
at the matter dominated epoch), $z_c$ (the redshift where equation of
state changes from $w_Q^m$ to $w_Q^0$), and $\Delta$ (the width of
transition). For the simulation we again use three sets of values in
order of increasing growth rate of $w(z)$ : (a) $w_Q^0=-0.8,
w_Q^m=-0.5, z_c^m=1.5, \Delta=0.1$, (b) $w_Q^0=-1.0, w_Q^m=-0.2,
z_c^m=0.6, \Delta=0.07$, and (c) $w_Q^0=-1.2, w_Q^m=0.1, z_c^m=0.11,
\Delta=0.03$.

We find that in both cases, the ansatz recovers the measured quantity
to within $0.5\%$ accuracy in the redshift range important for SNe
observations. Thus we find that even for fits for which the ansatz
does not return exact values, it can recover cosmological quantities
to a high degree of accuracy.


\begin{thebibliography}{}
\bibitem[\protect\citeauthoryear{Alam \&
    Sahni}{2002}]{as02}
Alam, U. and Sahni, V., 2002, {\tt astro-ph/0209443}.
\bibitem[\protect\citeauthoryear{Alam, Sahni \& Starobinsky}{2003}]{alam03}
Alam, U., Sahni, V. and Starobinsky, A.A., 2003, JCAP {\bf 0304}, 002, [{\tt astro-ph/0302302}].
\bibitem[\protect\citeauthoryear{Alam \etal}{2003}]{alam03a}
Alam, U., Sahni, V., Saini, T. D., and Starobinsky, A.A., 2003, \mn,
{\bf 344}, 1057
[{\tt astro-ph/0303009}].
\bibitem[\protect\citeauthoryear{Alcaniz, Jain and Dev}{2002}]{alcaniz}
Alcaniz, J.S., Jain. D. and Dev, A., 2002, \pd {\bf 66}, 067301
[{\tt astro-ph/0206448}].
\bibitem[\protect\citeauthoryear{Amendola}{2000}]{amendola00}
Amendola, L., 2000, \pd {\bf 62}, 043511.
\bibitem[\protect\citeauthoryear{Barber \etal}{2000}]{barber00}
Barber, A.J., \etal, 2000, \apj {\bf 545}, 444.
\bibitem[\protect\citeauthoryear{Barris \etal}{2003}]{hzt_new}
Barris, B. J. \etal, 2004, \apj {\bf 602}, 571B [{\tt astro-ph/0310843}].
\bibitem[\protect\citeauthoryear{Bartolo \& Pietroni}{2000}]{bp00}
Bartolo, N. and Pietroni, M. 2000 \pd {\bf 61}, 023518.
\bibitem[\protect\citeauthoryear{Bassett \etal}{2002}]{bassett}
Bassett, B.A., Kunz, M., Silk, J. and Ungarelli, C., 2002,
MNRAS, {\bf 336}, 1217 [{\tt astro-ph/0203383}].
\bibitem[\protect\citeauthoryear{Bertolami \& Martins}{2000}]{bm00}
Bertolami, O. and Martins, P.J., 2000, \pd {\bf 61}, 064007.
\bibitem[\protect\citeauthoryear{Birrell \& Davies}{1982}]{bd82}
Birrell, N.D. \& Davies, P.C.W., 1982, {\em Quantum Fields in Curved Space},
Cambridge University Press, Cambridge.
\bibitem[\protect\citeauthoryear{Boisseau \etal}{2000}]{beps00}
Boisseau, B., Esposito-Farese, G., Polarski, D. and Starobinsky, A.A., 2000, \prl {\bf 85}, 2236
\bibitem[\protect\citeauthoryear{Caldwell, Dave \&
    Steinhardt}{1998}]{cald98}
Caldwell, R.R., Dave, R. and  Steinhardt, P.J., 1998,
\prl {\bf 80}, 1582.
\bibitem[\protect\citeauthoryear{Caldwell}{2002}]{caldwell}
Caldwell, R.R., 2002, \plb {\bf 545}, 23 [{\tt astro-ph/9908168}].
\bibitem[\protect\citeauthoryear{Caldwell, Kamionkowski \& Weinberg}{2003}]{caldwell03}
Caldwell, R.R., Kamionkowski, M. and Weinberg, N.N., 2003,
Phys.Rev.Lett. {\bf 91} 071301 [{\tt astro-ph/0302506}].
\bibitem[\protect\citeauthoryear{Carroll}{2001}]{carroll01}
Carroll, S.M., 2001, Living Rev.Rel. {\bf 4} 1 [{\tt astro-ph/0004075}].
\bibitem[\protect\citeauthoryear{Carroll, Hoffman \& Trodden}{2003}]{carroll03}
Carroll, S.M., Hoffman, M. and Trodden, M., 2003, \pd {\bf 68}, 023509
[{\tt astro-ph/0301273}].
\bibitem[\protect\citeauthoryear{Cayrel \etal}{2001}]{cayrel01}
Cayrel, R. \etal, 2001, Nature {\bf 409}, 691 [{\tt astro-ph/0104357}].
\bibitem[\protect\citeauthoryear{Chaboyer \& Krauss}{2002}]{krauss02}
Chaboyer, B. and Krauss, L.M., 2002, ApJ, {\bf 567}, L45.
\bibitem[\protect\citeauthoryear{Chiba \&
    Nakamura}{2000}]{chiba}
Chiba, T. and Nakamura, T., 2000, \pd {\bf 62}, 121301(R).
\bibitem[\protect\citeauthoryear{Chiba, Okabe, \& Yamaguchi}{2000}]{chiba00}
Chiba, T., Okabe, T. and Yamaguchi, M, 2000, \pd {\bf 62}, 023511.
\bibitem[\protect\citeauthoryear{Chimento \etal}{2003}]{chimento03}
Chimento, L.P., Jakubi, A.A., Pavon, D. and Zimdahl, W., 2003, \pd {\bf 67}
083513 [{\tt astro-ph/0303145}].
\bibitem[\protect\citeauthoryear{Copeland, Liddle \& Lidsey}{2001}]{copeland01}
Copeland, E.J., Liddle, A.R. and Lidsey, J.E., 2001, \pd {\bf 64} 023509.
\bibitem[\protect\citeauthoryear{Corasaniti \&
    Copeland}{2003}]{copeland}
Corasaniti, P.S. and Copeland, E.J., 2003, \pd {\bf 67} 063521 [{\tt astro-ph/0205544}].
\bibitem[\protect\citeauthoryear{Corasaniti \etal}{2003}]{corasaniti}
Corasaniti, P.S., Bassett, B.A., Ungarelli, C. and Copeland, E.J., 2003,
\prl {\bf 90}, 091303 [{\tt astro-ph/0210209}].
\bibitem[\protect\citeauthoryear{Daly \& Djorgovsky}{2003}]{daly03}
Daly, R.A. and Djorgovsky, S.G., 2003, \apj {\bf 597}, 9-20.
[{\tt astro-ph/0305197}].
\bibitem[\protect\citeauthoryear{Damour, Kogan \& Papazoglou}{2002}]{damour02}
Damour, T., Kogan, I.I. and Papazoglou, A., 2002, \pd {\bf 66},
104025 [{\tt hep-th/0206044}].
\bibitem[\protect\citeauthoryear{Deffayet, Dvali \&
    Gabadadze}{2002}]{DDG}
Deffayet, C., Dvali, G. and Gabadadze, G., 2002, \pd {\bf 65},
044023 [{\tt astro-ph/0105068}].
\bibitem[\protect\citeauthoryear{Felder \etal}{2002}]{felder02}
Felder, G.N., Frolov, A., Kofman, L. and Linde, A., 2002,
Phys. Rev {\bf 66}, 023507.
\bibitem[\protect\citeauthoryear{Frampton}{2003}]{frampton03a}
Frampton, P., 2003, \plb {\bf 555}, 139.
\bibitem[\protect\citeauthoryear{Frampton \& Takahashi}{2003}]{frampton03b}
Frampton, P. and Takahashi, T., 2003, \plb {\bf 557}, 135.
\bibitem[\protect\citeauthoryear{Freedman \etal}{2001}]{HST}
Freedman, W., \etal, 2001, \apj {\bf 553}, 47.
\bibitem[\protect\citeauthoryear{Gerke \&
    Efstathiou}{2002}]{efstathiou}
Gerke, B, \& Efstathiou, G., 2002, \mn {\bf 335} 33,
[{\tt astro-ph/0201336}].
\bibitem[\protect\citeauthoryear{Gnedin, Lahav \& Rees}{2001}]{gnedin01}
Gnedin, N, Lahav, O. and Rees, M.J., {\tt astro-ph/0108034}.
\bibitem[\protect\citeauthoryear{Gratton \etal}{2003}]{gratton03}
Gratton, R. \etal, 2003, A\& A, {\bf 408}, 529, [{\tt astro-ph/0307016}].
\bibitem[\protect\citeauthoryear{Grib, Mamaev \& Mostepanenko}{1980}]{gmm80}
Grib, A.A., Mamaev, S.G. and Mostepanenko, V.M., 1980, {\em Quantum
Effects in Strong External Fields}, Moscow, Atomizdat (in Russian)
[English translation:
{\em Vacuum Quantum Effects in Strong Fields}, Friedmann Laboratory
Publishing, St.Petersburg, 1994].
\bibitem[\protect\citeauthoryear{Hansen \etal}{2002}]{hansen02}
Hansen, B. \etal, 2002, ApJ {\bf 574}, L155.
\bibitem[\protect\citeauthoryear{Hawking \& Ellis}{1973}]{hawellis}
Hawking, S.W. and Ellis, G.F.R., 1973, {\em The large scale structure of
space-time}, Cambridge University Press.
\bibitem[\protect\citeauthoryear{Hoffman}{2003}]{hoffman03}
Hoffman, M., 2003, {\tt astro-ph/0307350}.
\bibitem[\protect\citeauthoryear{Huterer \&
    Starkman}{2002}]{hut03}
Huterer, D. and Starkman, G., 2003, \prl {\bf 90}, 031301,
[{\tt astro-ph/0207517}].
\bibitem[\protect\citeauthoryear{Huterer \&
    Turner}{1999}]{turner}
Huterer, D. and Turner, M.S., 1999, \pd,  {\bf 60}, 081301.
\bibitem[\protect\citeauthoryear{Johri}{2003}]{johri}
Johri, V.B., 2003, {\tt astro-ph/0311293}.
\bibitem[\protect\citeauthoryear{Kallosh \etal}{2002}]{kallosh02}
Kallosh, R., Linde, A., Prokushkin, S. and Shmakova, M., 2002,
\pd {\bf 66} 123503.
\bibitem[\protect\citeauthoryear{Kamenshchik, Moschella \&
    Pasquier}{2001}]{chap1}
Kamenshchik, A., Moschella, U. and Pasquier, V., 2001, \plb {\bf 511} 265.
\bibitem[\protect\citeauthoryear{Knop \etal}{2003}]{knop}
Knop, R.A., et al., 2003, \apj {\bf 598} 102 [{\tt astro-ph/0309368}].
\bibitem[\protect\citeauthoryear{Krauss}{2001}]{krauss01a}
Krauss, L.M., 2001, in Proceedings, ESO-CERN-ESA Symposium on Astronomy, Cosmology and Fundamental Physics, March 2002.
\bibitem[\protect\citeauthoryear{Krauss}{2001}]{krauss01}
Krauss, L.M., 2001, in International Conference on the identification of
Dark Matter, Eds. N. Spooner and V. Kudryavtsev (Singapore, World Scientific), 1.
\bibitem[\protect\citeauthoryear{Krauss \& Chaboyer}{2001}]{krauss01b}
Krauss, L.M. and Chaboyer, B., 2001, {\tt astro-ph/0111597}.
\bibitem[\protect\citeauthoryear{Kunz \etal}{2003}]{kunz03}
Kunz,  M., Corasaniti, P., Parkinson, D.  and Copeland, E.J., 2003, {\tt astro-ph/0307346}.
\bibitem[\protect\citeauthoryear{Linder}{2003}]{linder}
Linder, E.V., 2003, \prl {\bf 90} 091301, [{\tt astro-ph/0208512}].
\bibitem[\protect\citeauthoryear{Maeda, Mizuno \& Torii}{2003}]{maeda}
Maeda, K., Mizuno, S. and Torii, T., 2003, \pd {\bf 68} 024033 [{\tt gr-qc/0303039}].
\bibitem[\protect\citeauthoryear{Maor \etal}{2002}]{maor02}
Maor, I. \etal, 2002, \pd {\bf 65} 123003, [{\tt astro-ph/0112526}].
\bibitem[\protect\citeauthoryear{Marchi \etal}{2003}]{marchi03}
Marchi, G. De, \etal, 2004, \asta {\bf 415}, 971 [{\tt astro-ph/0310646}].
\bibitem[\protect\citeauthoryear{McInnes}{2002}]{innes}
McInnes, B., 2002, JHEP {\bf 0208}, 029 [{\tt hep-th/0112066}].
\bibitem[\protect\citeauthoryear{Nakamura \&
    Chiba}{1999}]{nak99}
Nakamura, T. and Chiba, T., 1999, \mn, {\bf 306}, 696.
\bibitem[\protect\citeauthoryear{Nunes \& Lidsey}{2003}]{nunes}
Nunes, N.J. and Lidsey, J.E., 2003, {\tt astro-ph/0310882}.
\bibitem[\protect\citeauthoryear{Padmanabhan}{2003}]{paddy03}
Padmanabhan, T., 2003, Phys. Rep. {\bf 380}, 235
[{\tt hep-th/0212290}].
\bibitem[\protect\citeauthoryear{Parker \&
    Raval}{1999}]{parker99}
Parker, L. and Raval, A., 1999, \pd {\bf 60}, 063512, 123502.
\bibitem[\protect\citeauthoryear{Peebles \& Ratra}{1988}]{peebles88}
Peebles, P.J.E. and Ratra, B., 1998, Ap. J. Lett. {\bf 325}, L17.
\bibitem[\protect\citeauthoryear{Peebles \& Ratra}{2002}]{pr02}
Peebles, P.J.E. and Ratra, B., 2002, Rev.Mod.Phys. {\bf 75}, 559
[{\tt astro-ph/0207347}].
\bibitem[\protect\citeauthoryear{Peebles \& Vilenkin}{1999}]{pv99}
Peebles, P.J.E. and Vilenkin, A., 1999 \pd {\bf 59} 063505.
\bibitem[\protect\citeauthoryear{Percival \etal}{2001}]{2dF}
Percival, W.J., \etal, 2001, \mn {\bf 327}, 1297.
\bibitem[\protect\citeauthoryear{Perlmutter \etal}{1999}]{perl}
Perlmutter, S.J., \etal, 1999, \apj {\bf 517}, 565.
\bibitem[\protect\citeauthoryear{Riess \etal}{1998}]{riess}
Riess, A.G., \etal, 1998, \aj {\bf 116}, 1009.
\bibitem[\protect\citeauthoryear{Sahni}{2002}]{sahni02a}
Sahni, V., 2002, Class. Quantum Grav. {\bf 19}, 3435, [{\tt astro-ph/0202076}].
\bibitem[\protect\citeauthoryear{Sahni \&
    Habib}{1998}]{sh98}
Sahni, V. and Habib, S., 1998,  \prl {\bf 81}, 1766,
[{\tt hep-ph/9808204}].
\bibitem[\protect\citeauthoryear{Sahni \etal}{2003}]{sahni03}
Sahni, V., Saini, T.D., Starobinsky, A.A. and Alam, U., 2003,
\jetpl {\bf 77} 201
[{\tt astro-ph/0201498}].
\bibitem[\protect\citeauthoryear{Sahni, Sami \& Souradeep}{2002}]{sahni02c}
Sahni, V., Sami, M. and Souradeep, T., 2002, \pd {\bf 65} 023518.
\bibitem[\protect\citeauthoryear{Sahni \&
    Shtanov}{2003}]{ss02a}
Sahni, V. and Shtanov, Yu.V., 2003, JCAP {\bf 11} 014, [{\tt astro-ph/0202346}].
\bibitem[\protect\citeauthoryear{Sahni \& Starobinsky}{2000}]{ss00}
Sahni, V. and Starobinsky, A.A., 2000, IJMP D {\bf 9}, 373
[{\tt astro-ph/9904398}].
\bibitem[\protect\citeauthoryear{Sahni \& Wang}{2000}]{sahwang}
Sahni, V. and Wang, L., 2000, \pd {\bf 62}, 103517 
[{\tt astro-ph/9910097}].
\bibitem[\protect\citeauthoryear{Saini}{2003}]{saini03}
Saini, T.D., 2003, \mn {\bf 344}, 129.
\bibitem[\protect\citeauthoryear{Saini \etal}{2000}]{saini00}
Saini, T.D., Raychaudhury, S., Sahni, V. and Starobinsky, A.A., 2000,
\prl, {\bf 85}, 1162.
\bibitem[\protect\citeauthoryear{Sheth \etal}{2003}]{sheth03}
Sheth, J.V., Sahni, V., Shandarin, S.F. and Sathyaprakash, B.S., 2003,
MNRAS {\bf 343}, 22 [{\tt astro-ph/0210136}].
\bibitem[\protect\citeauthoryear{Singh, Sami \& Dadhich}{2003}]{singh}
Singh, P., Sami, M. and Dadhich, N.K., 2003, \pd {\bf 68}, 023522
[{\tt hep-th/0305110}].
\bibitem[\protect\citeauthoryear{Spergel \etal}{2003}]{spergel03}
Spergel, D.N. \etal, 2003,
\apj Suppl. {\bf 148}, 175 [{\tt astro-ph/0210136}]
\bibitem[\protect\citeauthoryear{Starobinsky}{1998}]{st00}
Starobinsky, A.A., 1998, \jetpl {\bf 68}, 757.
\bibitem[\protect\citeauthoryear{Tegmark}{2002}]{teg02}
Tegmark, M., 2002, \pd {\bf 66}, 103507. 
\bibitem[\protect\citeauthoryear{Tegmark \etal}{2003}]{tegmark}
Tegmark, M., \etal 2003, {\tt astro-ph/0310723}.
\bibitem[\protect\citeauthoryear{Thompson \etal}{2001}]{thompson01}
Thomson, I.B. \etal, 2001, \aj {\bf 121}, 3089.
\bibitem[\protect\citeauthoryear{Tonry \etal}{2003}]{tonry03}
Tonry, J.L., et al., 2003, \apj {\bf 594}, 1, [{\tt astro-ph/0305008}].
\bibitem[\protect\citeauthoryear{Urena-Lopez \&
    Matos}{2000}]{um00}
Urena-Lopez, L.A., Matos, T., 2000, \pd {\bf 62}, 081302,
[{\tt astro-ph/0003364}].
\bibitem[\protect\citeauthoryear{Wang \& Lovelace}{2001}]{wang01}
Wang, Y and Lovelace, G, 2001, \apj {\bf 562}, L115.
\bibitem[\protect\citeauthoryear{Weinberg}{1989}]{wein89}
Weinberg, S. (1989) Rev. Mod. Phys. {\bf 61}, 1.
\bibitem[\protect\citeauthoryear{Weller \& Albrecht}{2002}]{albrecht}
Weller, J. and Albrecht, A., 2002, \pd {\bf 65}, 103512
[{\tt astro-ph/0106079}].
\bibitem[\protect\citeauthoryear{Wetterich}{1988}]{wetterich88}
Wetterich, C, 1988, \nucp {\bf B302}, 668.
\bibitem[\protect\citeauthoryear{Zeldovich}{1968}]{zel68}
Zeldovich, Ya.B. (1968) Sov. Phys. -- Uspekhi {\bf 11}, 381.

\end{thebibliography}
\end{document}